\documentclass[
    prfluids,
    reprint, 
    onecolumn, 
    superscriptaddress,
    aps,
    nofootinbib
]{revtex4-2}

\usepackage[T1]{fontenc}
\usepackage{graphicx}
\usepackage{dcolumn}
\usepackage{bm}
\usepackage{float}
\usepackage{amsmath,amssymb,amsfonts}
\usepackage{todonotes}
\usepackage{epstopdf}
\usepackage{multirow}
\usepackage{booktabs}
\usepackage{enumitem}
\usepackage{comment}
\usepackage{overpic}

\usepackage{setspace}
\usepackage{esint}

\usepackage{color}   
\definecolor{darkGreen}{rgb}{0,0.45,0}
\definecolor{darkBlue}{rgb}{0,0,0.7}
\definecolor{darkRed}{rgb}{0.76, 0.13, 0.28}

\usepackage{hyperref}
\hypersetup{
    colorlinks=true,
    linktoc=all,
    linkcolor=darkBlue,
    citecolor=darkGreen,
    urlcolor=darkRed,
}

\begin{document}

\title{Nonlinear Aerodynamic Response and an Equivalent Static Wind-resistant Design for Anticlastic Conical Tensile Membranes}

\author{Ajay Kumar}%
\affiliation{Graduate Associate, Dept. of Civil Engg., Indian Institute of Technology Kanpur, UP 208016, India}

\author{Budhaditya De}%
\affiliation{Graduate Associate, Dept. of Civil and Environmental Engg., University of California, Los Angeles, USA}

\author{Sudib Kumar Mishra}%
\email[Electronic address: ]{smishra@iitk.ac.in}
\affiliation{Professor, Dept. of Civil Engg, Indian Institute of Technology Kanpur, UP 208016, India}

\author{Devasmit Dutta}
\affiliation{Graduate Associate, Dept. of Civil and Environmental Engg., University of California, Los Angeles, USA}

\date{\today}

\begin{abstract}
Conical Tensile Membrane Structure (TMS) is commonly used for aesthetics, economic design, high rain and snow loading. Such TMS shows complex aerodynamic behavior in presence of geometric nonlinearity, not adequately studied in the past. The aerodynamic responses of anticlastic conical TMS under random wind loading is presented herein along with an equivalent static wind resistant design approach. The stochastic wind loading on the TMS in the atmospheric boundary layer (ABL) is simulated via the Large Eddy simulation (LES); which is detailed in a previous study by the authors and hence not repeated here. The aerodynamic loading is then employed as input in conducting the nonlinear time history analyses considering open (i.e. without facade) and closed (with facade) TMS, supported by peripheral/radial cables. The influence of the key parameters (aerodynamic roughness height, the rise-span ratio of the TMS and the membrane prestress, notably) are demonstrated.  Although increasing prestress and rise-to-span ratio enhances the stiffness of TMS, the former shows dominance. Increasing roughness height also lead to increased peak loading/responses by enhanced turbulence. An equivalent static wind-resistant design is presented via the Gust Response Factors (GRFs) and an additional Nonlinear Adjustment Factors (NAFs). These factors are presented systematically, encompassing alternative scenarios. Multi-linear regression models are presented for predictive modeling of these factors, along with a probabilistic analysis for their design values that can be employed in practice bypassing an involved nonlinear dynamic analysis.
\end{abstract}

\maketitle

\noindent\textbf{Keywords:}
Tensile membrane structures (TMS), Large Eddy Simulation (LES),
nonlinear dynamic analysis, Gust Response Factor,
Nonlinear Adjustment Factors

\section{Introduction}

Structural engineering community has largely embraced the utilization of the Tensile membrane structures (TMSs) in the recent past for their ability to span large span of areas for community centers, airport, shopping complexes, sporting stadiums and so on. The TMS also offer aesthetically pleasing geometry by their characteristic synclastic or anticlastic curvatures \cite{otto1973tensile}. The TMS are exemplified by the iconic Denver International Airport terminal roof, the Khan Shatyr entertainment center in Kazakhstan, the Olympic stadium roofs in Montreal and Berlin and the Millennium Dome in London, to name a few. They embody the perfect fusion of form and function. Appeal of TMs also lies in their efficient load carrying mechanism, leading to lightweight nature and minimal usage of material. However, the slender and flexible nature of TMS also induces vulnerability to long period dynamic excitations, such as wind \cite{gosling2013analysis,rizzo2018peak,kumar2025aerodynamic,kumar2025aerodynamica}. Although several standards have been formulated over the years \cite{american2017minimum,american2016tensile,forster2004introduction,cecs2004technical} for assessing the design wind loading on the TMS, there is still the need for a widely acceptable robust wind-resistant design philosophy for such structures. Thus, investigations need to be carried out to gain insight into the incoming wind flow field and complex nonlinear wind-structure interactions in the TMS, especially if they are doubly curved. Past attempt in these directions is reviewed herein to pinpoint the specific novelty of this study.\\

Prior to the prevalence of the nonlinear analysis of highly deformable structures, Uematsu and Uchiyama \cite{uematsu1982wind} conducted aeroelastic wind tunnel tests (WTTs) on a pretensioned suspended roof model. The WTT on a variety of TMS geometries including mono-sloped canopies, the Expo Boulevard (a long-span TMS) in Shanghai, China \cite{zhou2013research}, ridge and valley type TMS \cite{sun2019investigation}, spherical inflatable type \cite{chen2022wind}, arch-supported oval-shaped \cite{sun2020investigation}, and saddle type \cite{kandel2022wind} were conducted in the recent past. Hincz and Gamboa-Marrufo demonstrated the impact of membrane deformation on the spatial distribution of pressure coefficients \cite{hincz2016deformed}. Pargana et al. \cite{pargana2010fully}, Dinh et al. \cite{dinh2015study,dinh2016numerical}, and Xu et al. \cite{xu2022experiment} conducted nonlinear finite element analyses of the tensioned surfaces to assess the dynamic response characteristics, validated through experiments. The dynamic properties of thin film membranes was investigated by numerical simulation and experiments by Young et al. \cite{young2005numerical}. Li and Chan conducted an analysis of flexible, frame-supported membranes \cite{li2004integrated}. Lee and Youn carried out nonlinear, large deformation finite element analysis of wrinkled membranes \cite{lee2006finite}. A recent study by Rizzo et al. employed a finite element aeroelastic membrane model for determining natural frequencies and aerodynamic damping subjected to varying incoming wind speed \cite{rizzo2021investigation}.\\

Equivalent static wind loading was proposed instead of dynamic wind by Kandel et al. \cite{kandel2021wind} for arch-supported oval shaped TMS and by Chen et al. \cite{chen2022wind} for spherical inflatable TMS based on WTT on scaled rigid model.\\

While standard practices rely on the WTT for obtaining/verifying the wind loading, computational fluid dynamic (CFD) simulation of flow field are also conducted recently to complement the WTT. The CFD has demonstrated remarkable efficacy in simulating complex flow fields around intricate structures. De Nayer et al. \cite{de2018numerical,de2022fsi} conducted the fluid-structure interaction (FSI) study between an air-inflated flexible hemispherical membrane and the surrounding wind flow using the Large Eddy Simulations (LES). The FSI study reveals the coupling between the wind flow and the membrane dynamics. Michalski et al. \cite{michalski2011validation} validated full-scale aerodynamic test results on a 29m flexible umbrella canopy through computational FSI to propose a virtual design methodology for lightweight flexible membranes under fluctuating wind. Vázquez \cite{vazquez2007nonlinear} analyzed orthotropic membranes and shells using nonlinear FSI simulations. Wind-induced vibrations of highly deformable membranes and shells were investigated by Gluck et al. \cite{gluck2003computation} using a partitioned coupling scheme for time-dependent FSI. Kupzok \cite{kupzok2009modeling} conducted a fully coupled FSI analysis with a partitioned solution strategy on “ARIES” mobile canopy, detailing the aeroelastic effects by the turbulent wind boundary layer.\\

Majority of the existing literature presents hyperbolic paraboloid or spherical geometries (for doubly curved forms) or mono-sloping cylindrical shapes. Despite being one of the most commonly used TMS designs, research on the anticlastic conical form under wind excitations remains scanty. Therefore, this research work aims to specifically assess the nonlinear aero-dynamic responses of anticlastic conical TMS under incoming wind flow. A variety of TMS geometry with pertinent parameters (rise-span ratio, membrane prestress and ground roughness) are considered. The Large Eddy simulation (LES) is employed for generations of the incident wind profile in the form of a homogeneous atmospheric boundary layer (ABL). Details of the LES has been presented in an earlier study by the authors \cite{de2023comparative,de2024large} and hence is not repeated herein. Nevertheless, the dynamic pressure time histories on the TMS surface are obtained as input in the nonlinear (finite element based) dynamic analysis for determining the dynamic responses, such as the out-of-plane displacements, in-plane principal stresses in the membrane and the axial tension in the cables supporting the TMS. Both the open and closed TMS configurations, supported by the peripheral and radial supporting cables are considered.\\

Furthermore, the outcome of the dynamic analysis is condensed into an equivalent static wind-resistant design methodology for facilitating the design. The gust loading factors, and a nonlinear adjustment factor are adopted for this purpose. A multi-linear regression model is fitted between peak aerodynamic responses and the aforementioned factors; thus, allowing the designers to estimate the peak responses directly bypassing the need for complex nonlinear dynamic analysis.

\section{Geometry of the anticlastic TMS}\label{sec:geometry-TMS-anticlastic}

Geometry of the anticlastic TMS adopted in this study is presented herein. The aerodynamic pressure field around the TMS is obtained via Computational Fluid Dynamic (CFD) simulation. The CFD make use of the LES to accurately capture the intricate flow features. The dynamic pressure time histories are obtained on the TMS surface to be used as input in conducting the nonlinear dynamic response analysis.\\

\begin{figure}[htbp!]
    \centering
    \hspace*{-.75cm}
    \includegraphics[width=.82\textwidth]{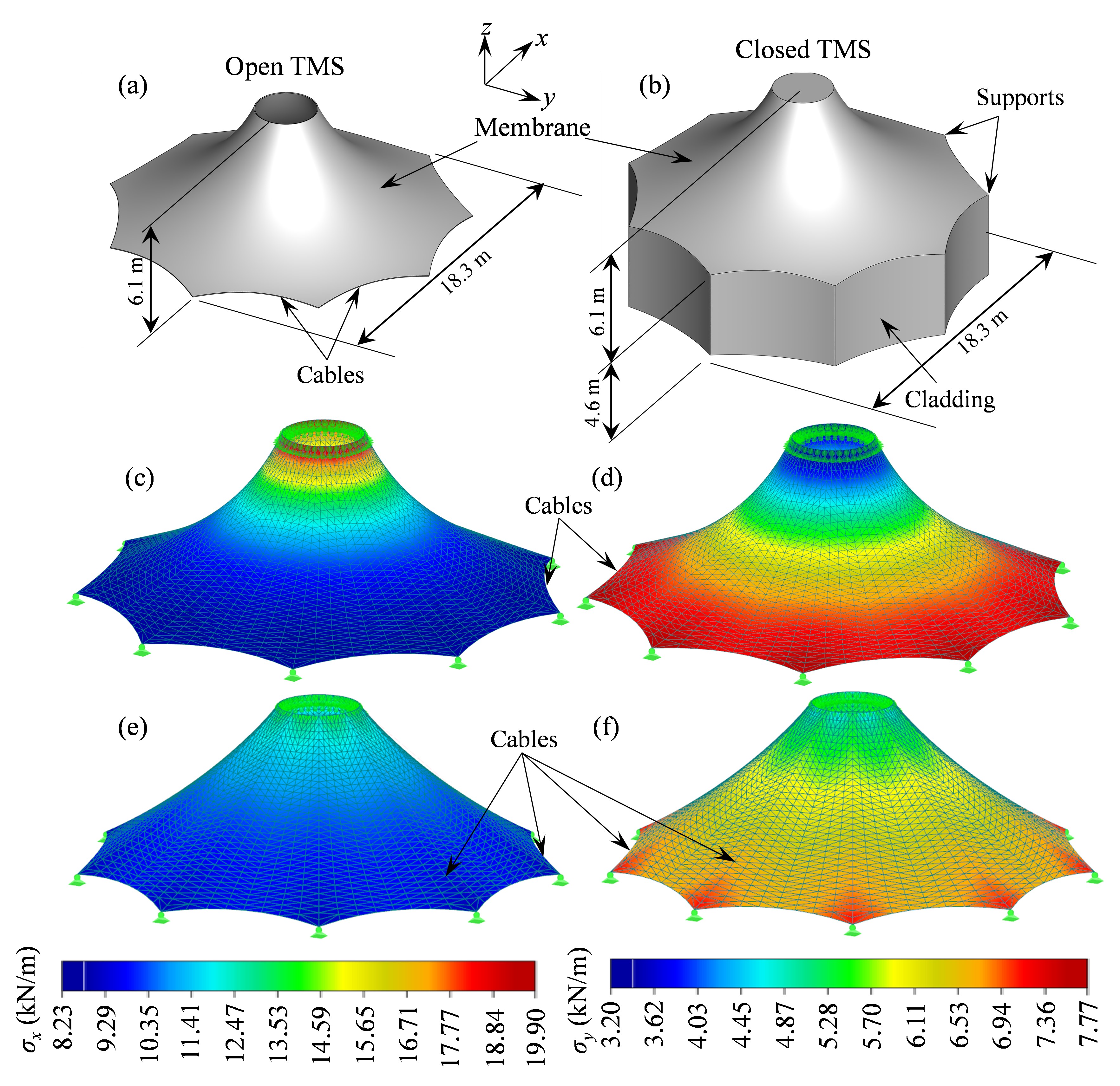}
    \caption{
    Model details for (a) Open TMS and (b) Closed TMS; Final form and distribution of prestress along (c) warp direction without radial cables; (d) weft direction without radial cables; (e) warp direction with radial cables; (f) weft direction with radial cables}
    \label{fig:modeldetails}
\end{figure}

A nine-sided, anticlastic conical TMS is adopted in this study, as shown in Fig.~\ref{fig:modeldetails}(a)-(d). The geometries for the open and closed configurations of the TMS are shown in Fig.~\ref{fig:modeldetails}(a) and Fig.~\ref{fig:modeldetails}(b), respectively. The membrane is supported by cables either (a) at the bottom periphery or (b) both along the bottom periphery as well as radially along the membrane surface, as in Fig.~\ref{fig:modeldetails}(c)-(d) and Fig.~\ref{fig:modeldetails}(e)-(f). At the top, the membrane is supported by a ring beam, framed into a central mast. The load from the cables is transferred to the supporting columns. The aerodynamic LES does not require modelling of the ring beam, central mast and the columns but the membrane surface. The influence of columns on wind loading on TMS is negligible due to its low solidity ratio~\cite{mara2014influence}. Furthermore, the dynamic response evaluation emphasizes the response of membrane and cables under dynamic wind loading. The top circular edge and cable end nodes are assumed to be simply supported in the dynamic analysis.\\

TMSs are force-active structures as their final configuration is governed by the applied prestress. The final form is obtained through a technique referred as “form-finding”. Out of several alternatives, the Updated References Strategy by Bletzinger and Ramm~\cite{bletzinger1999general} is used to obtain the membrane geometry for a given level of prestress. The final configuration and the distribution of prestress (along warp and weft directions) of the conical membrane are shown in Fig.~\ref{fig:modeldetails}(c)-(d) for TMS without radial cables and Fig.~\ref{fig:modeldetails}(e)-(f) for TMS with both radial and peripheral cables. Note that for the example shown, the TMS has a span $(L)$ of 18.3 m, rise-span ratio $(f/L)$ of 1/3 eaves height $(h)$ of 4.6 m and prestress $(N_0)$ of 8 kN/m.\\

Material properties of the membrane and cable materials are listed in Table~\ref{tab:material_properties}. In order to determine the influence of key parameters (namely the rise-span ratio, membrane prestress and aerodynamic roughness height) on the wind-induced response of the TMS, both the LES for the wind loading and the dynamic analysis for the responses are performed for different models. The set of parameters considered for this study are given in Table~\ref{tab:membrane_parameters}.\\

\begin{table}[h!]
\centering
\caption{Material properties for the membrane and the cables}
\label{tab:material_properties}
\begin{tabular}{l l l l}
\hline
\textbf{Membrane Properties} &  & \textbf{Cable Properties} &  \\
\hline
Elastic modulus ($E$)   & 550 MPa   & Elastic modulus ($E$)   & 200 GPa \\
Thickness ($t$)         & 2 mm      & Diameter ($d$)          & 12 mm   \\
Shear modulus ($G$)     & 209 MPa   & Poisson’s ratio ($\nu$) & 0.30    \\
Poisson’s ratio ($\nu$) & 0.314     & Density ($\rho$)        & 7850 kg/m$^3$ \\
Density ($\rho$)        & 2250 kg/m$^3$ &                     &          \\
\hline
\end{tabular}
\end{table}

\begin{table}[h!]
\centering
\caption{Membrane geometric and aerodynamic loading parameters}
\label{tab:membrane_parameters}
\begin{tabular}{l l}
\hline
\textbf{Parameter} & \textbf{Values} \\
\hline
Rise-to-span ratio ($f/L$)              & $1/6,\; 1/3,\; 1/2$ \\
Membrane prestress ($N_0$)              & $4\,\text{kN/m},\; 8\,\text{kN/m},\; 15\,\text{kN/m}$ \\
Span ($L$)                              & $18.3\,\text{m}$ \\
Eaves height ($h$)                     & $4.6\,\text{m}$ \\
Aerodynamic roughness height ($z_0$)   & $0.1\,\text{cm},\; 6\,\text{cm},\; 80\,\text{cm}$ \\
\hline
\end{tabular}
\end{table}

\section{Large Eddy simulation (LES) of wind loading on the membrane}\label{sec:les-membrane}

Time histories of the dynamic wind loading on the membrane are evaluated using the LES. Details of the LES, governing equations, simplifications and assumptions are well presented in the existing literature and hence not repeated herein for brevity. A detailed description of the LES on anticlastic, conical TMS may be obtained from the previous work of the authors~\cite{de2023comparative,de2024large}.\\

\begin{figure}[htbp!]
    \centering
    \hspace*{-.75cm}
    \includegraphics[width=.82\textwidth]{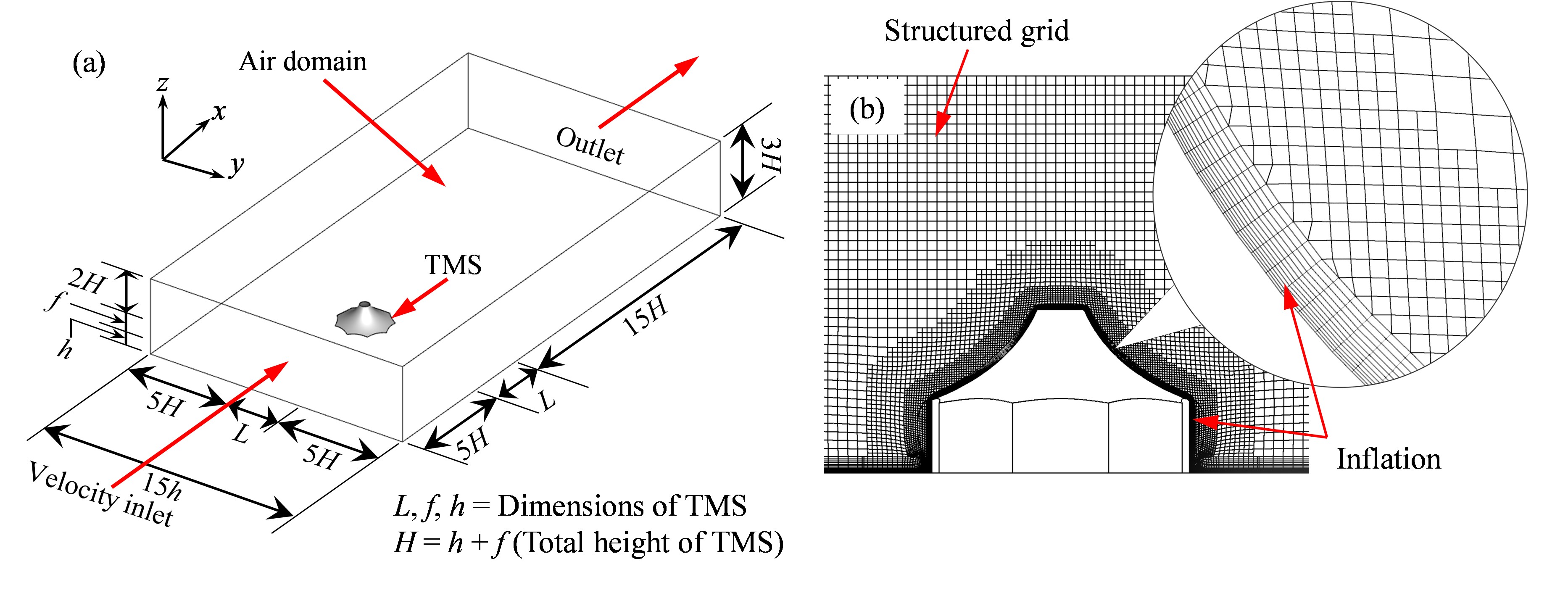}
    \caption{
    (a) Computational domain and the (b) Finite volume (FV) meshing around the closed TMS}
    \label{fig:computationaldomain}
\end{figure}

The dimensions of the rectangular computational domain, shown in Fig.~\ref{fig:computationaldomain}(a), are chosen following the guidelines by Franke et al.~\cite{franke2007introduction}. The inlet, lateral, top, and outlet boundaries are positioned at sufficient distances from the TMS by ensuring noninterference. These dimensions result in a blocking ratio of 2.745 percent, which is below the maximum specified ratio of 3 percent~\cite{dagnew2013computational,thordal2019review}. The fluid domain is discretized using the cut-cell meshing technique~\cite{fidkowski2007triangular,iousef2017use}. Different grid sizes ($f/120$, $f/12$, and $f/10$) are employed at the membrane surface, ground and the bulk of the domain, respectively. Additionally, an inflation growth factor of 1.10 is used to gradually increase the grid size within the domain. The discretized fluid domain surrounding the closed TMS is shown on Fig.~\ref{fig:computationaldomain}(b) The LES\textsubscript{IQ} criterion is employed to verify the accuracy of the chosen mesh size~\cite{celik2005index}, yielding an LES\textsubscript{IQ} criterion of 88 percent, exceeding the minimum prescribed value of 85 percent~\cite{gousseau2013quality,zheng2020cfd}.\\

A horizontally homogeneous and neutrally stratified Atmospheric boundary layer (ABL) profile is applied as the velocity inlet boundary condition. The applied profiles of mean wind velocity, $u_x$ (along flow direction $x$), turbulence kinetic energy, $k$ and turbulence dissipation rate, $\varepsilon$ are given as~\cite{richards1993appropriate,guichard2019assessment,blocken2007cfd}:

\begin{equation}
\label{eq:e1}
u_x(z) = \frac{u_*}{\kappa} \ln\!\left( \frac{z + z_0}{z_0} \right)
\end{equation}

\begin{equation}
\label{eq:e2}
k(z) = \frac{3}{2} \left( u_x(z)\, I(z) \right)^2
\end{equation}

\begin{equation}
\label{eq:e3}
\varepsilon(z) = \frac{u_*^3}{\kappa \left( z + z_0 \right)}
\end{equation}

where $\kappa$ (= 0.4) is the von Kármán's constant, $z_0$ is the aerodynamic roughness height (governed by terrain category), $u_*$ is the wall function friction velocity (calculated using ~\eqref{eq:e1} by substituting $u_x$ with $U_{\mathrm{ref}}$, which is the reference flow velocity at a reference height $z_{\mathrm{ref}}$). For the present study $U_{\mathrm{ref}}$ is considered as 50 m/s at a reference height $z_{\mathrm{ref}}$ of 10 m. The symbol $I$ in ~\eqref{eq:e2} is the turbulence intensity, the profile of which is given by:

\begin{equation}
\label{eq:e4}
I(z) = \frac{A}{\ln\!\left( \dfrac{z + z_0}{z_0} \right)}
\end{equation}

In this expression, constant $A$ is determined by substituting $I$ with $I_{\mathrm{ref}}$, i.e., the reference turbulence intensity at the reference height $z_{\mathrm{ref}}$. This study considers $I_{\mathrm{ref}}$ as 10, 20 and 35 percent for $z_0$ of 0.1 cm, 6 cm and 80 cm, respectively based on the terrain category~\cite{american2012wind}. Symmetry boundary conditions are applied at the upper and lateral (side) boundaries, which implies zero normal velocity and zero normal gradients of all the variables. The outlet boundary is assigned as pressure-outlet. Both the ground and membrane surfaces are assigned as smooth wall boundaries. \\

The commercial code Ansys Fluent v21.1 is used for all CFD simulations~\cite{fluent2021ansys}. The portion of the unresolved turbulence spectrum in the LES requires a Sub-Grid Scale model based on filtering of the Navier-Stokes equations. For this unresolved part of the turbulence spectrum, the Wall-Adapting Local Eddy-Viscosity (WALE) formulation proposed by Nicoud and Ducros~\cite{nicoud1999subgrid} is applied herein. \\

A second-order scheme and a bounded central-difference scheme are employed for pressure and momentum, respectively, for spatial discretization. A fractional-step pressure-velocity coupling, a Non-Iterative Time Advancement (NITA) scheme is applied for the temporal discretization. The ANSYS Fluent Theory Guide provides specifics on the associated algorithms~\cite{fluent2021ansys}. To strictly adhere to the CFL (Courant-Friedrichs-Lewy) condition in every cell, a fixed timestep of $5 \times 10^{-4}$ is chosen. The flow statistics are computed over thirty flow-through times, after six flow-through times of the LES initialization run.

\section{Verification of the LES outcome}\label{sec:verification-les}

\begin{figure}[htbp!]
    \centering
    \hspace*{-.75cm}
    \includegraphics[width=.82\textwidth]{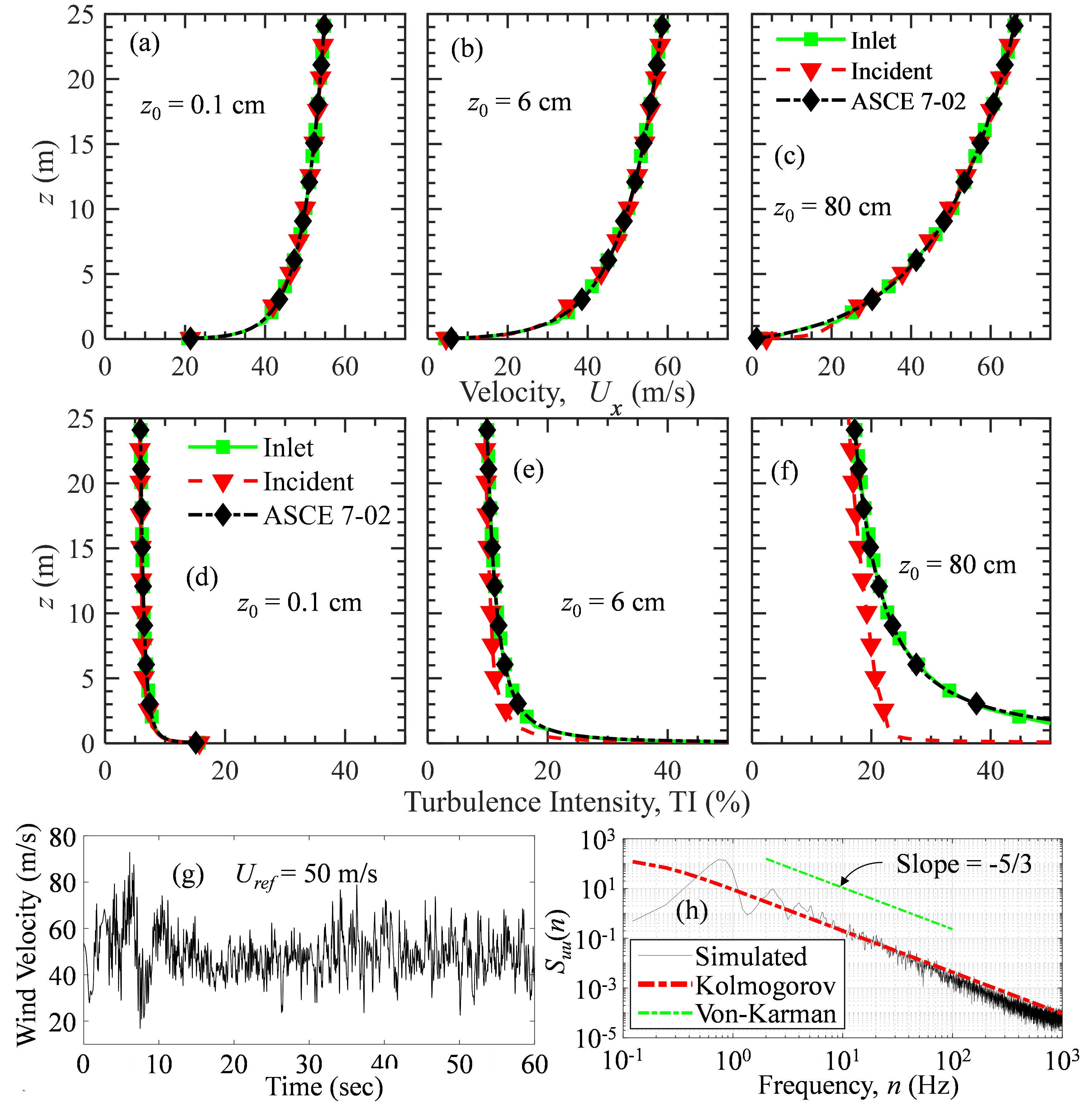}
    \caption{
    Longitudinal mean flow velocity profiles for   (a) 0.1, (b) 6 cm and (c) 80 cm. Respective turbulence intensities in (d), (e) and (f); (g) Time history of wind velocity recorded at   near the point of incidence to the TMS. (h) Simulated wind velocity PSD and the Von-Karman spectrum at   near the point of incidence to the TMS}
    \label{fig:lesverification}
\end{figure}

The simulations are initially carried out in an empty domain to check the feasibility of boundary layer profiles and whether they are in line with those prescribed in relevant standards. Fig.~\ref{fig:lesverification}(a)-(c) show the height-wise mean longitudinal wind velocity for three distinct roughness heights considered ($z_0$ = 0.1 cm, $z_0$ = 6 cm, $z_0$ = 80cm). Variation of the longitudinal turbulence intensity along the height for this roughness are also shown in Fig.~\ref{fig:lesverification}(d)-(f). The simulated profiles are plotted at the inlet and point of incidence with the TMS and compared with the ones prescribed in ASCE 7-16~\cite{american2017minimum}. These profiles show clear correlation between the simulated profiles and the ASCE suggested one, with fully developed ABL maintaining its longitudinal homogeneity along the flow direction.\\

For a specific roughness height ($z_0$= 80 cm), the longitudinal wind velocity fluctuations at a location on the reference height (i.e. 10 m above the ground) and just before the incidence to the membrane are shown in Fig.~\ref{fig:lesverification}(g). Respective power-spectral density (PSD) is shown in Fig.~\ref{fig:lesverification}(h). The von Kármán velocity spectrum and the simulated PSDs are seen to be in good agreement.\\

A larger disparity between the inlet and incident turbulence profiles, close to the ground surface, is noted for higher roughness heights (e.g., $z_0$ = 80 cm). This is because large roughness acts as a distinct obstacle to hinder the flow, causing low-pressure recirculation zones and vortex shedding downstream, thereby disturbing the three-dimensional isotropic turbulence structure by preventing the flow mixing and ultimately suppressing the turbulence intensity. Low $z_0$, however does not hinder the flow enough to stop such flow mixing and associated processes, and the inlet/incident profiles remain mostly same, and the flow/turbulence structure remains relatively uniform/unaffected.\\

The efficacy of the LES, the flow modeling is initially tested on a hyperbolic paraboloid rigid roof from Colliers et al.~\cite{colliers2020mean}. The validation of the process has been presented in detail by the authors in~\cite{de2024large}. Time histories of the wind pressure on the membrane surface (assuming the membrane to be rigid) are obtained from the LES. The pressure time histories, so obtained are subsequently employed on the TMS to conduct the dynamic response analysis. Such analysis, therefore, does not account for the coupling between the wind flow field and the dynamic membrane response (and vice versa); which is rather accounted in a detailed Fluid-Structure Interaction (FSI) study, not considered herein. 

\section{Nonlinear dynamic analysis of the TMS under simulated wind loading}\label{sec:non-linear-dyn-analysis-tms}

Wind pressure time histories obtained from the LES on specific locations on the membrane are applied as random excitation. The nonlinear finite element, time domain analysis is carried out for evaluation of dynamic responses. The commercial finite element code ABAQUS 2022 is used for all the dynamic analyses~\cite{manual2022abaqus}. Geometric nonlinearity is also considered. The wind pressure time histories are applied to act normally to the instantaneous configuration of the deformed membrane like a follower force. The influence of key parameters (rise-span ratio, $f/L$; aerodynamic roughness, $z_0$, membrane pretension, $N_0$) are then studied systematically on the aerodynamic responses. It is reemphasized that the analysis does not account for aero-elastic coupling between the membrane and wind loading.\\

\begin{figure}[htbp!]
    \centering
    \hspace*{-.75cm}
    \includegraphics[width=.82\textwidth]{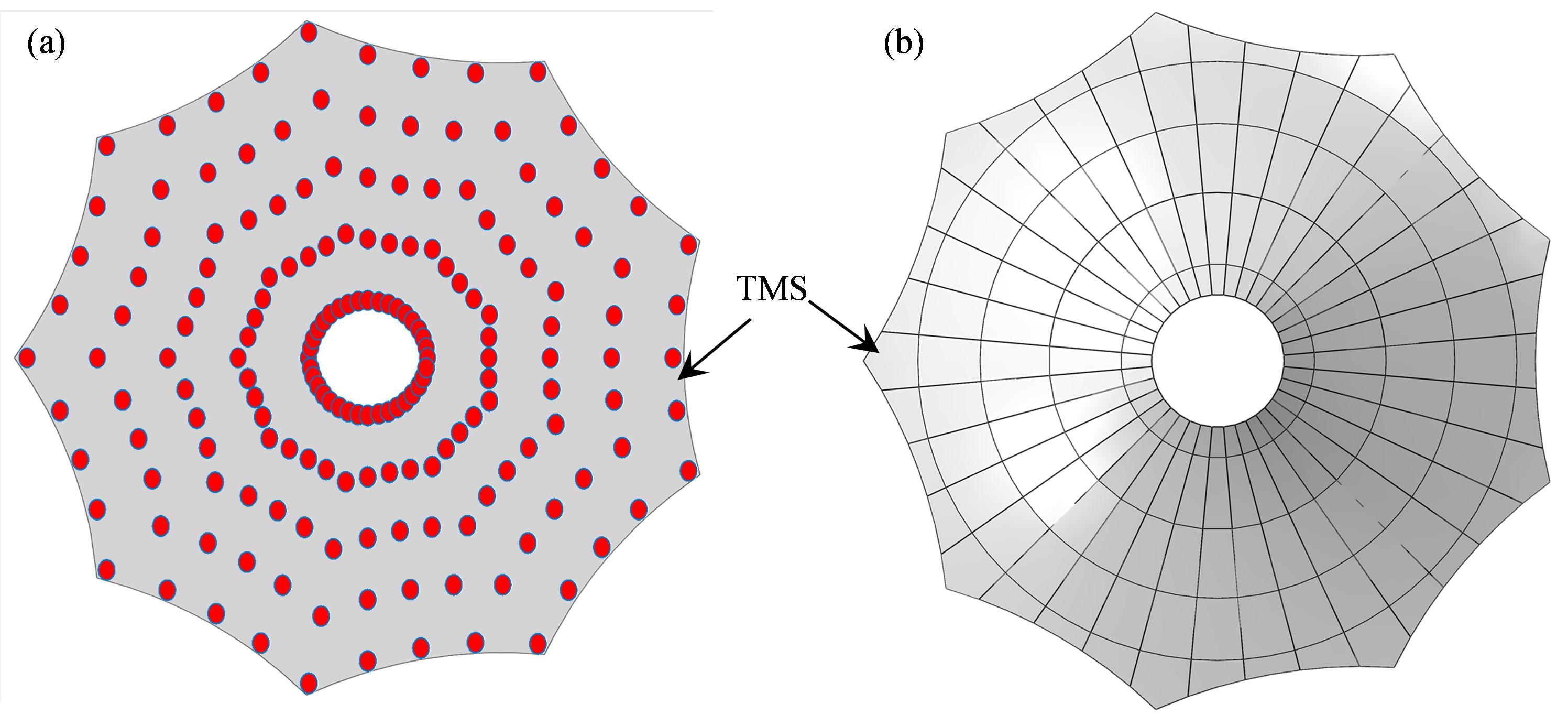}
    \caption{
    (a) Points on the TMS surface and (b) corresponding tributary area of each point monitored for recording pressure time histories}
    \label{fig:figure4}
\end{figure}

Fig.~\ref{fig:figure4} shows the points on the TMS surface for sampling of the pressure time histories. The membrane surface is partitioned into radial lines at $10^{0}$ apart, on each of these 5 points are taken along the height; generating a total of 180 sampling points as shown in Fig.~\ref{fig:figure4}. The membrane is modeled using finite elements and is discretized with 4-noded quadrilateral membrane elements with 3 translational degrees of freedom per node. The cables are modeled using the 2-noded 3-dimensional truss elements with 3 degrees of freedom per node.\\

The equation of motion for the membrane under uncoupled aerodynamic loading is given as:

\begin{equation}
\label{eq:e5}
[M]\{\ddot{U}\} + [C]\{\dot{U}\} + [K(\{U\})]\{U\} = \{F\}
\end{equation}

Where $[M]$ is mass matrix and $[C]$ is the damping matrix. The stiffness matrix $[K]$ is a function of the displacement ${U}$ due to geometric nonlinearity. The vectors $\{U\}$, $\{\dot{U}\}$, and $\{\ddot{U}\}$ are the displacement, velocity and acceleration, respectively; $\{F\}$ is a vector of external forces. The damping matrix is obtained following Rayleigh damping with 2 percent viscous damping, which is in the range for membrane following Forster et al.~\cite{forster2004introduction}. Eq.~\eqref{eq:e5} can be rewritten as

\begin{equation}
\label{eq:e6}
[M]\{\ddot{U}\} + \{P(\{U\},\{\dot{U}\})\} = \{F\}
\end{equation}

The load vector $\{P\}$ $(\{P\} = [C]\{\dot{U}\} + [K(\{U\})]\{U\})$ is taken as the vector of internal forcing in the membrane. Nonlinear dynamic response of the TMS under wind loading is determined by utilizing the Newmark implicit average acceleration time integration scheme to ensure unconditional stability. The equation of motion at $(n+1)^{\text{th}}$ time step can be written as:

\begin{equation}
\label{eq:e7}
[M]\{\ddot{U}\}_{n+1} + \{P(\{U\},\{\dot{U}\})\}_{n+1} = \{F\}_{n+1}
\end{equation}

The acceleration and velocity vectors an $(n+1)^{\text{th}}$ time step can be written as per Newmark’s average acceleration method as:

\begin{equation}
\label{eq:e8}
\{\ddot{U}\}_{n+1}
=
C_0 \left( \{U\}_{n+1} - \{U\}_n \right)
- C_1 \{\dot{U}\}_n
- \{\ddot{U}\}_n
\end{equation}

\begin{equation}
\label{eq:e9}
\{\dot{U}\}_{n+1}
= C_2(\{U\}_{n+1} - \{U\}_n)
- \{\dot{U}\}_n
\end{equation}

Where, $C_0 = \frac{4}{\Delta t^{2}}$, $C_1 = \frac{4}{\Delta t}$ and $C_2 = \frac{2}{\Delta t}$. $\Delta t$ is the time step for integration. The subscript $n$ in the above equations corresponds to the quantities at $n^{\text{th}}$ time step. Substituting Eq.~\eqref{eq:e8} and Eq.~\eqref{eq:e9} into Eq.~\eqref{eq:e7}, the following form is obtained as:

\begin{equation}
\label{eq:e10}
C_0[M]\{U\}_{n+1}
+ \{P(\{U\},\{\dot{U}\})\}_{n+1}
= \{F\}_{n+1}
+ [M]\left( C_0\{U\}_n + C_1\{\dot{U}\}_n + \{\ddot{U}\}_n \right)
\end{equation}

At $n^{\text{th}}$ time, step, all quantities in the right side of above equation are known. Eq.~\eqref{eq:e10} is a nonlinear as $\{P\}_{n+1}$ is a function of $\{U\}_{n+1}$. An iterative algorithm based on the Newton-Raphson technique is adopted to converge at the solution for each step. For this, Eq.~\eqref{eq:e10} is expressed in terms of residual forces or error function $\{R\}_{n+1}$ as:

\begin{equation}
\label{eq:e11}
\{R\}_{n+1}
= \{F\}_{n+1}
+ [M]\left( C_0\{U\}_n + C_1\{\dot{U}\}_n + \{\ddot{U}\}_n \right)
- C_0[M]\{U\}_{n+1}
- \{P\}_{n+1}
\end{equation}

If $\{U\}_{n+1}^{\,i}$ is and approximate trial solution at the $i^{\text{th}}$ iteration producing an error $\{R\}_{n+1}^{\,i}$, an improved solution $\{R\}_{n+1}^{\,i+1}$ is obtained using first order Taylor expansion of $\{R\}_{n+1}^{\,i+1}$ and by equating to zero as:

\begin{equation}
\label{eq:e12}
\{R\}_{n+1}^{\,i+1}
\approx
\{R\}_{n+1}^{\,i}
+ [K_T]_{n+1}^{\,i}
\left( \{U\}_{n+1}^{\,i+1} - \{U\}_{n+1}^{\,i} \right)
= \{0\}
\end{equation}

Where, $[K_T]_{n+1}^{\,i}$ is the tangent stiffness matrix. Eq.~\eqref{eq:e12} is written in incremental form as:

\begin{equation}
\label{eq:e13}
[K_T]_{n+1}^{\,i}\{\Delta U\}_{n+1}^{\,i}
= -\{R\}_{n+1}^{\,i}
\end{equation}

Where, $\{\Delta U\}_{n+1}^{\,i}$ is the displacement increment in the current iteration as:

\begin{equation}
\label{eq:e14}
\{\Delta U\}_{n+1}^{\,i}
= \{U\}_{n+1}^{\,i+1} - \{U\}_{n+1}^{\,i}
\end{equation}

Improved solution at $(i+1)^{\text{th}}$ iteration is obtained as:

\begin{equation}
\label{eq:e15}
\{U\}_{n+1}^{\,i+1}
= \{U\}_{n+1}^{\,i} + \{\Delta U\}_{n+1}^{\,i}
\end{equation}

The above iterative procedure is carried out until the error function $\{R\}_{n+1}^{\,i+1}$ is sufficiently close to zero to get the response at $(n+1)^{\text{th}}$ time step. This procedure is repeated for all the successive time steps until convergence of the scheme.

\section{Results and discussion}\label{sec:results}

This section presents the results and discussion on the nonlinear aerodynamic responses of the conical membrane. The analysis begins with the results from modal analysis, highlighting the deformed pattern of the initial mode shapes and examining the influence of key parameters, namely the rise-span ratio ($f/L$) and the membrane prestress ($N_0$), on the natural frequencies of the TMS. Subsequently, the impact of these critical parameters along with the aerodynamic roughness height ($z_0$) is analysed on the wind-induced responses.   

\subsection{Modal Analysis}

\begin{figure}[htbp!]
    \centering
    \hspace*{-.75cm}
    \includegraphics[width=.82\textwidth]{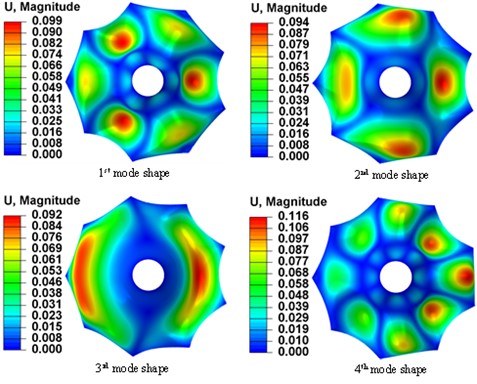}
    \caption{
    Mode shapes of TMS with peripheral cables only and rise-span ratio $(f/L)$ = 1/3 and membrane prestress $(N0)$ 8 kN/m   }
    \label{fig:figure5}
\end{figure}

The natural frequencies and the sets of mode shapes are obtained through the modal analysis. Fig.~\ref{fig:figure5} illustrates the first four mode shapes of the TMS without radial cables and a rise-span ratio ($f/L$) of 1/3 and a membrane prestress ($N_0$) of 8 kN/m. Fig.~\ref{fig:figure5} depicts the impact of rise-span ratio and membrane prestress on the natural frequencies of the TMS. The model frequencies are seen to increase with increasing rise-span ratio and the membrane prestress. The rate of frequency increase is more pronounced for membrane prestress compared to the rise-span ratio, which is because prestress have a more pronounced influence on the membrane stiffness by directly increasing the geometric stiffness. Together, higher prestress and a greater rise-span ratio both contribute to increased structural stiffness. Additionally, the rate of increase in natural frequencies with respect to the rise-span ratio decreases for higher modes.	 \\

\begin{figure}[htbp!]
    \centering
    \hspace*{-.75cm}
    \includegraphics[width=.82\textwidth]{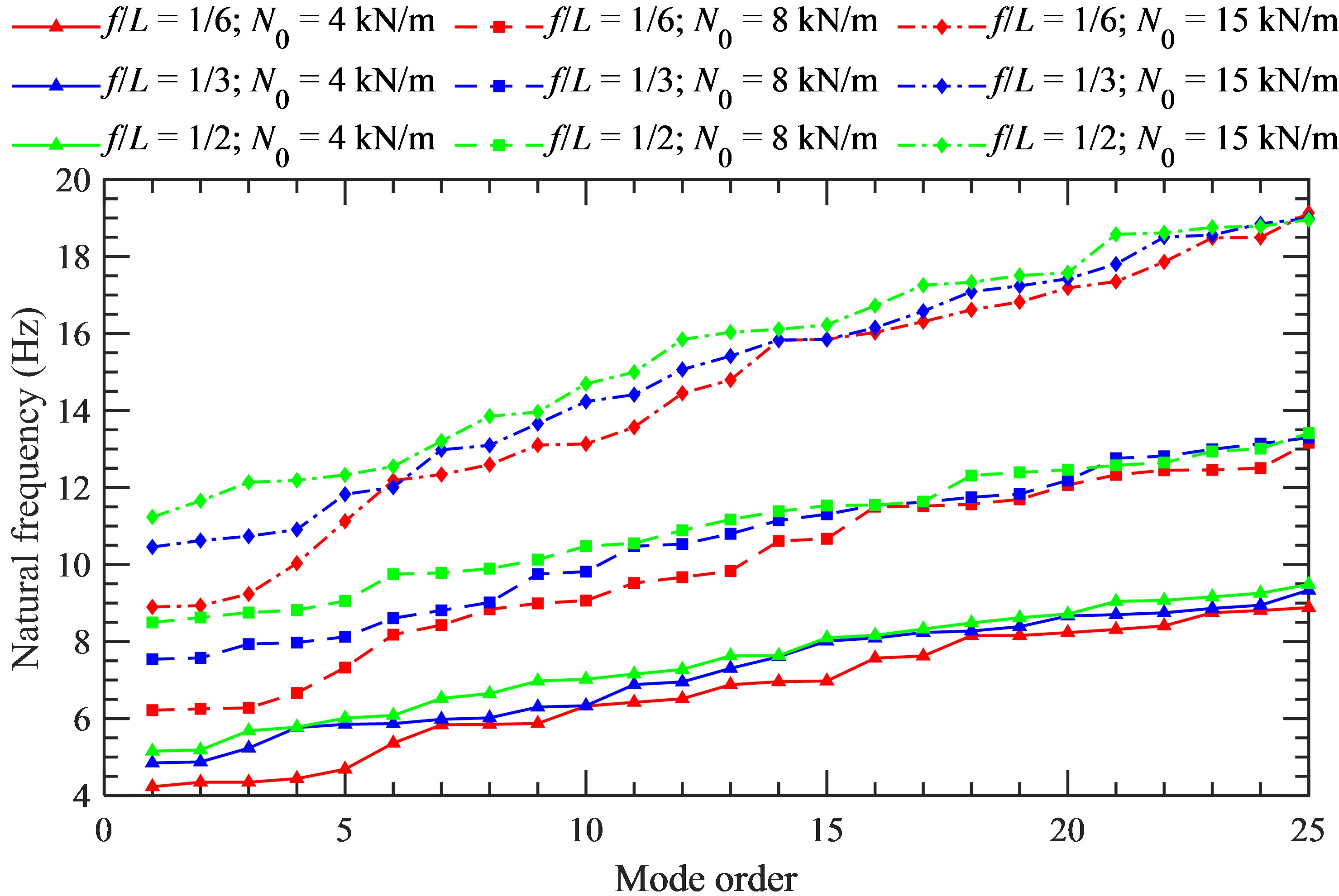}
    \caption{
    Modal frequencies with increasing rise-to-span ratio and membrane prestress for TMS with peripheral cables only   }
    \label{fig:figure6}
\end{figure}

Following the symmetry of the TMS, the mode shapes also display varying folds of symmetry and anti-symmetry. The folds of symmetry/anti-symmetry do not necessarily increase with increasing modal frequencies. For instance, the first mode shows a threefold of symmetry-antisymmetry, second mode shows two-folds of symmetry-antisymmetry and third mode shows merely one fold of symmetry and so on. Such symmetry also led to identical frequencies for different modes, as may be seen in Fig. \ref{fig:figure6}. The modes are also coupled due to geometric nonlinearity, allowing modal interactions to give rise to complex dynamic responses, explored in the subsequent sections. Influence of the key structural parameters, ($f/L$) ratio and membrane prestress ($N_0$) individually on each of the output responses are systematically presented in the following subsections. 

\subsection{Wind-induced responses of the conical TMS }
\begin{figure}[htbp!]
    \centering
    \hspace*{-.75cm}
    \includegraphics[width=.82\textwidth]{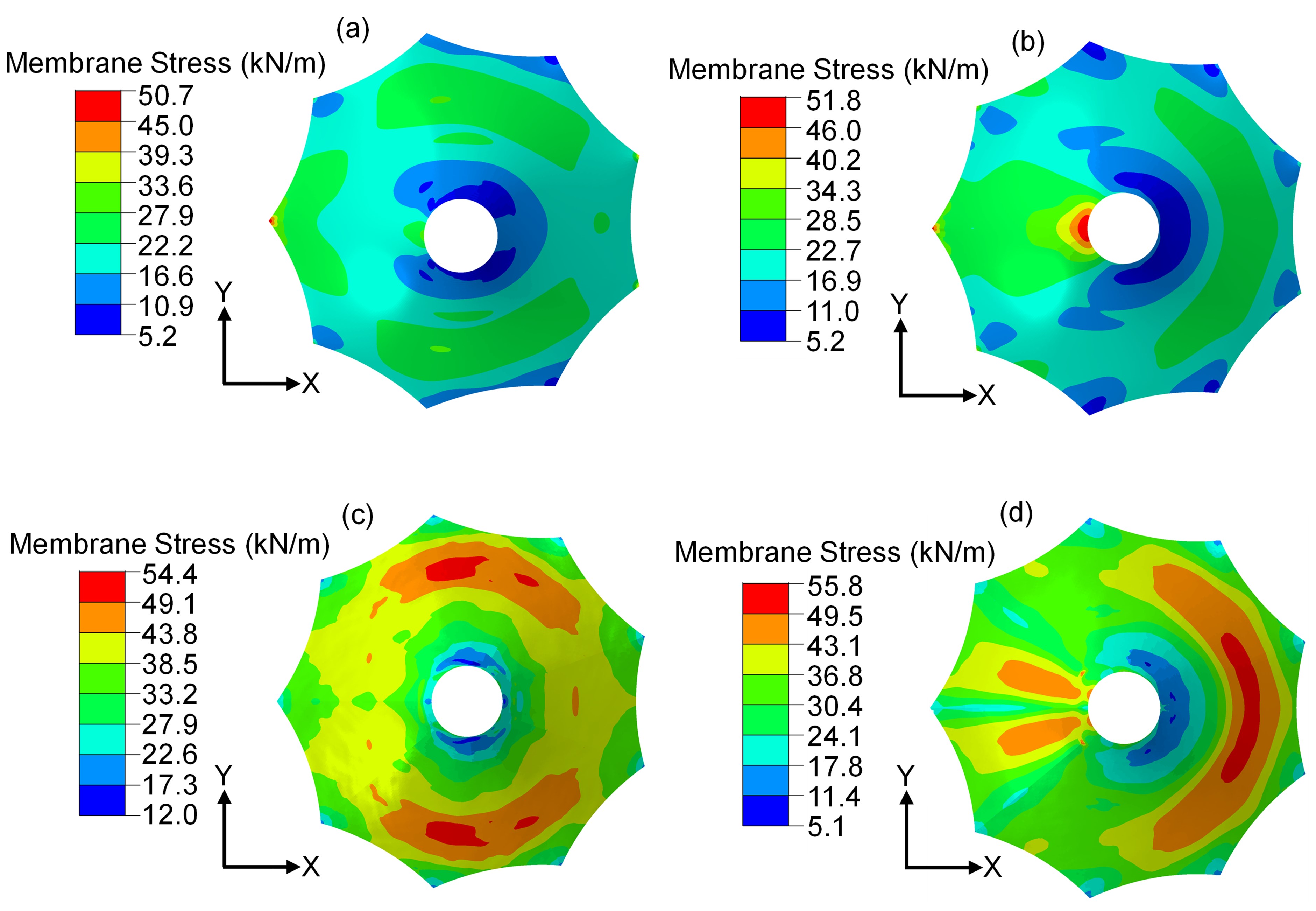}
    \caption{
    Principal stress contour in TMS with $f/L = 1/3$, $N_0 = 15 kN/m$ and $z_0 = 0.1 cm$ for (a) Closed TMS with only peripheral cables (b) Open TMS with only peripheral cables (c) Closed TMS with radial as well as peripheral cables (d) Open TMS with radial as well as peripheral cables }
    \label{fig:figure7}
\end{figure}

The spatio-temporal distribution of dynamic responses for alternative TMS configurations is obtained along with the influence of key parameters on the membrane displacements and principal stresses.  The maximum principal stress contours are shown in Fig. \ref{fig:figure7}.a-\ref{fig:figure7}.d. The time instant at which the maximum principal stress occurs on the membrane surface is taken for the snapshots. A particular example is shown for open and closed type membranes with and without radial cables for comparisons. Broadly, TMS with both radial and peripheral cables experience greater in-plane stresses by a factor of around 1.08. Further the distribution of high stress regions (depicted in yellow and above) are more widespread for TMS with radial and peripheral cables. The cables thus help in dispersing the stress, otherwise gets concentrated/banded around particular point/zone of the membrane.      \\

The open TMS experiences slightly higher stress than the closed one, which may be due to the combined contribution of pressure at the top and bottom surfaces. Most of the high stress regions for the open TMS occur on the leeward portion, experiencing net outward suction by the recirculating vortices/wake on the top and outward thrust of the recirculating wind from the bottom (De et al., \cite{de2024large}). The additive nature of both forcing leads to greater in-plane stresses. For the closed TMS, since there is only one surface, high stress regions occur near the flow separation regions on either side perpendicular to the wind flow direction. \\

\begin{figure}[htbp!]
    \centering
    \hspace*{-.75cm}
    \includegraphics[width=.82\textwidth]{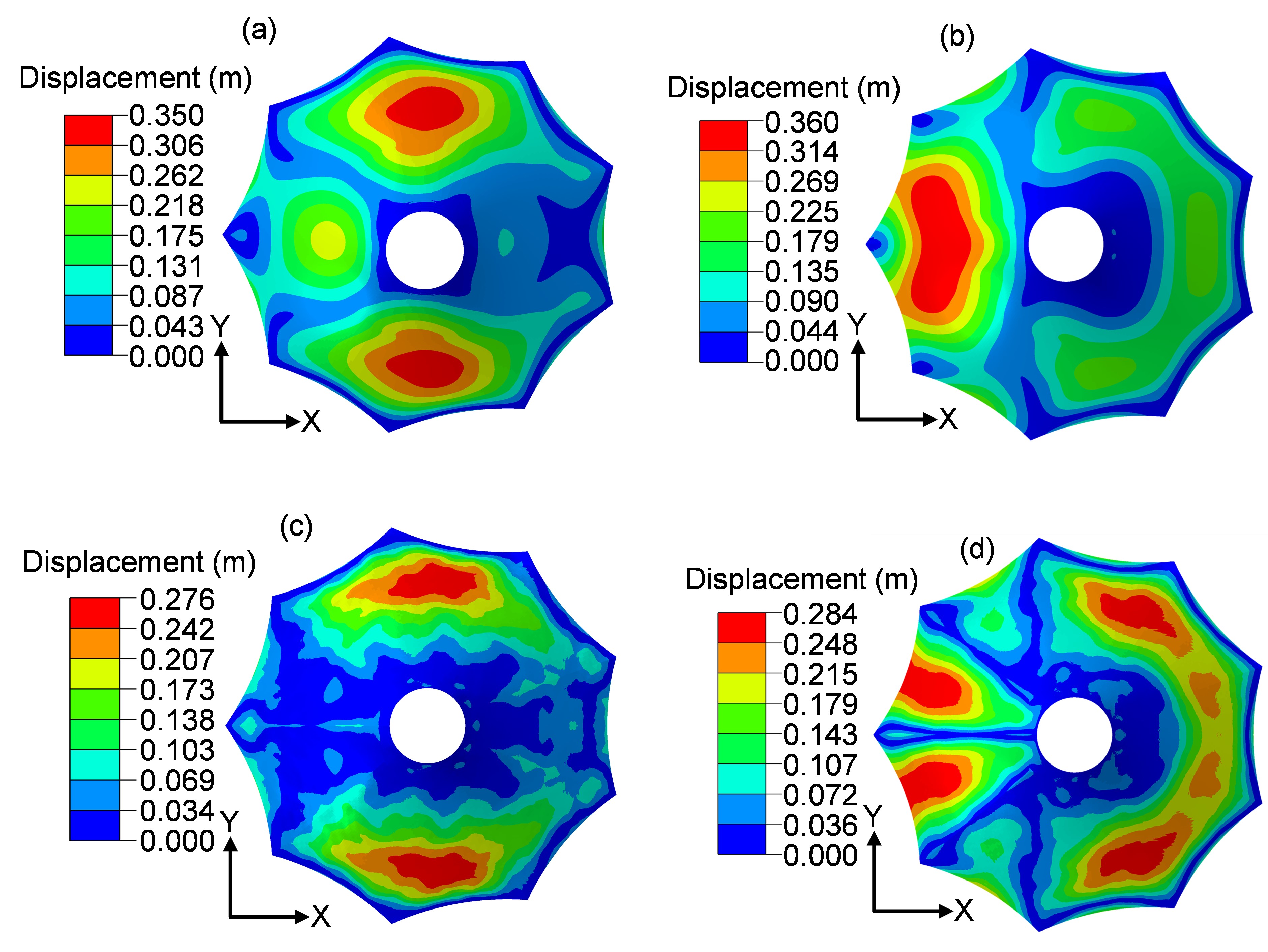}
    \caption{
    Deflection contour of membrane for TMS with $f/L = 1/3$, $N0 = 15 kN/m$ and $z_0 = 0.1 cm$ for  (a) Closed TMS with only peripheral cables (b) Open TMS with only peripheral cables (c) Closed TMS with radial and peripheral cables (d) Open TMS with radial and peripheral cables }
    \label{fig:figure8}
\end{figure}

In next, the out-of-plane deformation/deflections are presented using Fig. \ref{fig:figure8}.a-\ref{fig:figure8}.d. Alike membrane stresses, these contours also correspond to the time instant at which any location on the TMS surface experienced highest deflection out of the entire time-history. Expectedly, the TMS with both the radial and peripheral cables show reduced deflection (by 22 percent) due to increased overall stiffness when contrasted with the TMS with only peripheral cables. The closed TMS experiences lower displacements compared to the open ones because of the combined effect of the inward thrust exerted by the wind on the top surface and suction at the bottom surface. Alike membrane stresses, the locations of the maximum displacements are almost identical. In the case of closed TMS, large outward suction around the flow separation zones on either side causes large outward displacements. In case of the open TMS, peak displacements are observed in the windward region where the incoming wind strikes directly onto the surface in combination with the suction from the vortices below. The Leeward part also experiences some outward displacements in the case of the TMS with radial and peripheral cables.\\

\begin{figure}[htbp!]
    \centering
    \hspace*{-.75cm}
    \includegraphics[width=.82\textwidth]{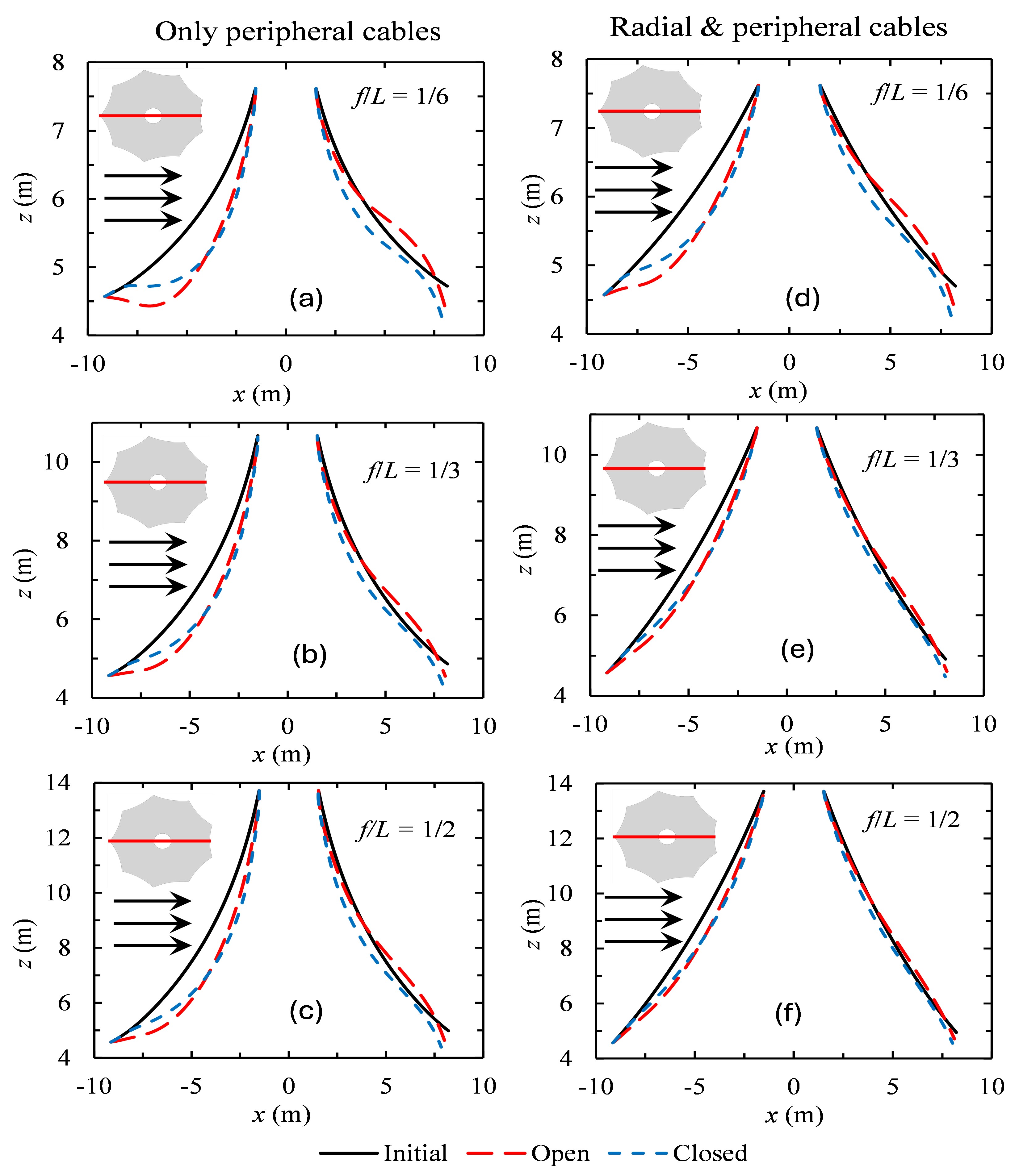}
    \caption{
    Vertical peak deformation profiles of the membrane for different $f/L$ ratios, $z_0 = 0.1 cm$, pretension of 15 kN/m and TMS with and without radial cables  }
    \label{fig:figure9}
\end{figure}

In order to exhibit the vertically deformed profiles of the TMS under peak deformation, sectional slices are contrasted along the plane of symmetry, along the flow direction in Fig. \ref{fig:figure9}.a-\ref{fig:figure9}.f. These confirm higher displacements for the open TMS and in TMS without radial cables. Furthermore, with the increasing $f/L$ ratio (especially from 1/6 to 1/3), the displacements are reduced to a certain extent. In parity with the modal analysis, higher $f/L$ leads to increased overall membrane stiffness and resulting in reduced displacements. The reduction, however, is modest when changing from $f/L = 1/3$ to $f/L =1/2$. 

\subsection{Effect of rise-span ($f/L$) ratio and prestress ($N_0$) on peak responses  }
The effect of the rise-to-span ratio and the membrane prestress are demonstrated on the peak dynamic responses. Due to their diverse stiffness attributes, the TMS with and without radial cables are presented separately in section 4.2.1 and 4.2.2, respectively.   

\subsection{TMS with only peripheral cables   }
The influence of rise-span ratio (f/L) on the peak membrane displacement, membrane principal stress, cable tensile force and reaction force at the supports are discussed herein. Fig. \ref{fig:figure10} shows the variation of these responses with varying $f/L$ ratio for distinct level of prestress (N0) and ground roughness (z0) for both open and closed type TMS. Increased stiffness with the increasing rise-to-span ratio is noted earlier. From a previous study by the authors (De at al., 2024), it is also observed that with increasing rise-span ratio, the mean pressure coefficients increase for open/closed TMS, indicating rising intensity of aerodynamic loading. The peak displacements, principal stresses and cable tension increase with increasing rise-to-span ratio, as increased ratio enhances both the membrane stiffness as well as the aerodynamic loading. This increase, however, is not uniform. For peak displacements and stress, a gradual rise is observed from $f/L$ = 1/6 to 1/3 followed by a sharp rise from $f/L$ = 1/3 to 1/2. The effect is more prominent for $N_0$ = 4 kN/m. \\

For a particular $f/L$, the peak displacements decrease with increasing prestress due to greater membrane stiffness. It also adds on to the peak principal stresses. Relatively less increase in peak cable tension is noted with increasing $f/L$ ratio. With $f/L$ ranging from 1/6 to 1/3, peak tension in the cables remains nearly identical for both types of TMS, irrespective of roughness heights (z0). With further increase in $f/L$ (from 1/3 to ½), increment in peak cable tension is observed for prestress $N_0$ = 4 kN/m and 8 kN/m. The trend for $N_0$ = 15 kN/m, contrarily becomes relatively regular for both TMS types with $z_0$ = 0.1 cm and 6 cm. For $z_0$ = 80 cm, peak cable tension first increases gradually from $f/L$ = 1/6 to 1/3 and then attains saturation for $f/L$ = 1/2. The effect of larger prestress attains prominence (especially for $N_0$ = 15 kN/m) as the peak cable tension becomes quite higher for both type of TMS, irrespective of $z_0$ values. \\
                    
The peak support reactions near the TMS base show the opposite trend to the rest of the responses. Common to all $N_0$ , the maximum support reactions decrease with rising $f/L$ ratio. With increasing rise-span ratio, the share of reaction by the central mast (at top circular edge of the TMS) increases to reduce the share of reaction on the pinned base support. The base reactions thus decrease with increasing rise-span ratio. The variation in reactions are prominent for prestress of 4 kN/m to 8 kN/m; whereas it increases nominally for $N_0$ = 15 kN/m. The open TMS show higher responses in most cases than the closed one, for reasons explained earlier. \\

 All responses show an increase with increasing roughness height since higher aerodynamic roughness causes higher peak factors for the wind pressure coefficients.\\

 \subsection{TMS with both radial as well as peripheral cables }
\begin{figure}[htbp!]
    \centering
    \hspace*{-.75cm}
    \includegraphics[width=.82\textwidth]{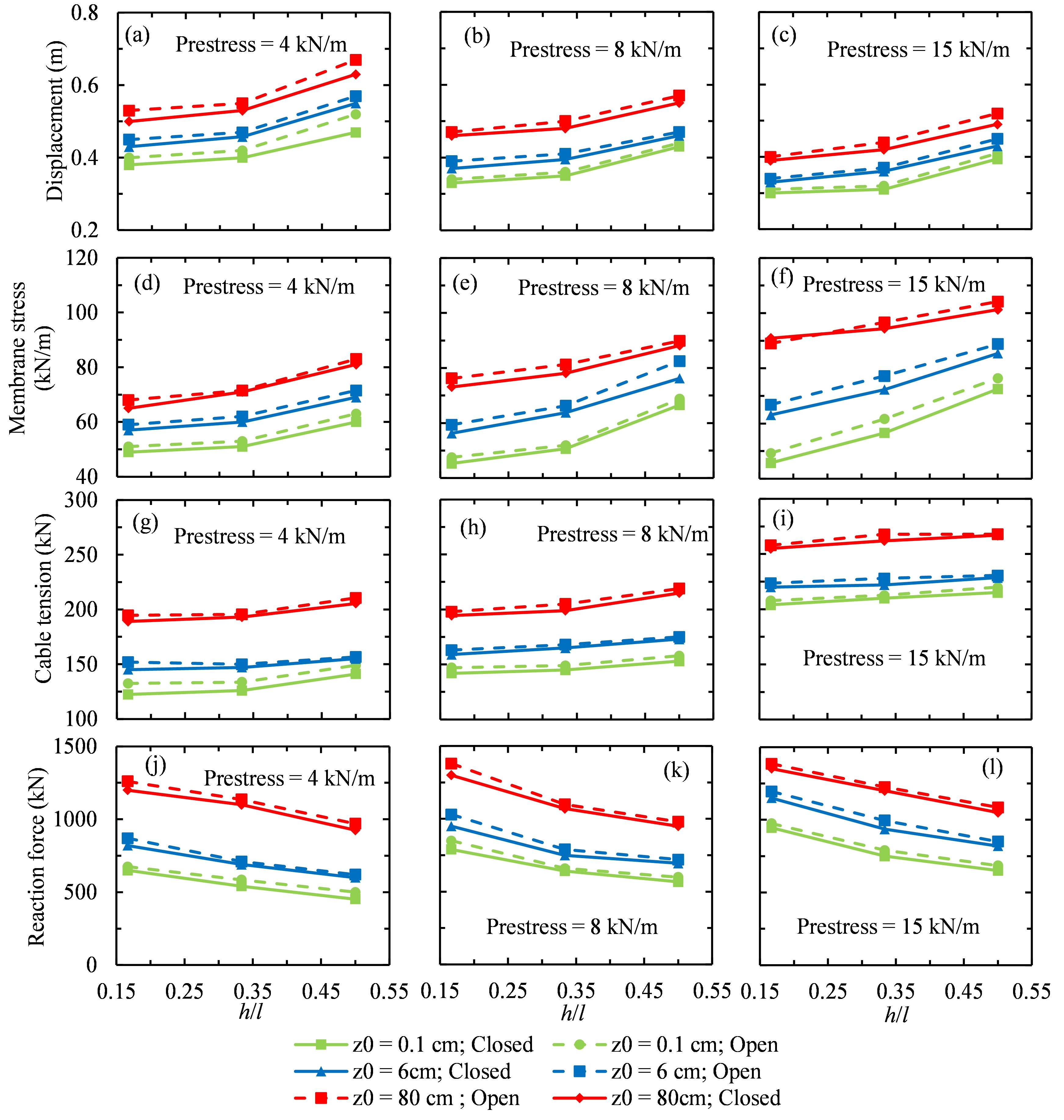}
    \caption{
    Influence of key parameters on the peak wind-induced responses for TMS with only peripheral cables, (a)-(c) Displacement, (d)-(f) Membrane stress, (g)-(i) Cable tensile force and (j)-(l) Support reaction force   }
    \label{fig:figure10}
\end{figure}

\begin{figure}[htbp!]
    \centering
    \hspace*{-.75cm}
    \includegraphics[width=.82\textwidth]{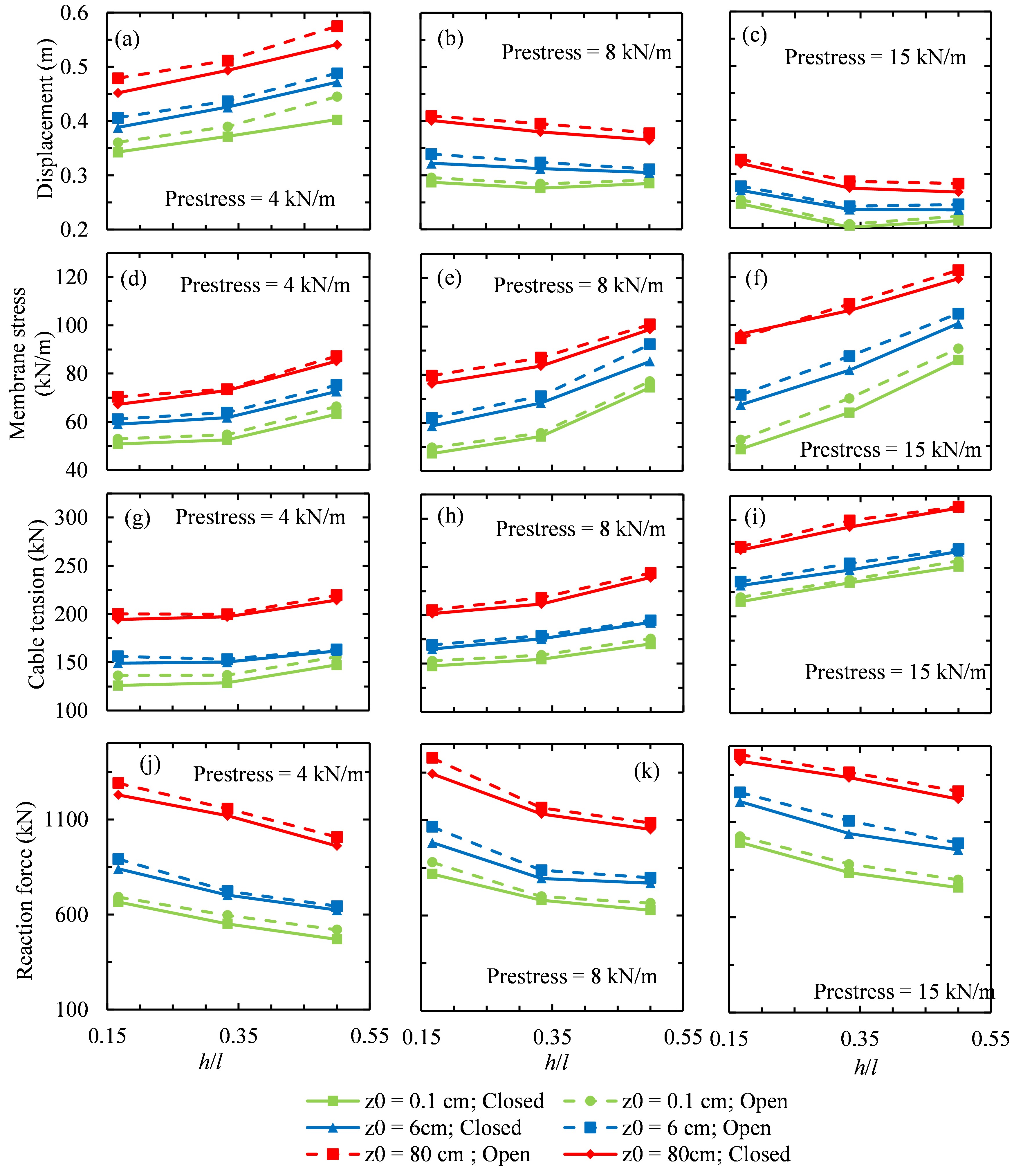}
    \caption{
    Influence of key parameters on the peak responses for TMS with radial and peripheral cables, (a)-(c) Displacement, (d)-(f) Membrane stress, (g)-(i) Cable tensile force and (j)-(l) Support reaction force   }
    \label{fig:figure11}
\end{figure}

 Pertinent response behavior for TMS with both radial and peripheral cables is presented in Fig. \ref{fig:figure11}.a-\ref{fig:figure11}.d. The structural responses follow a similar trend as in Fig. \ref{fig:figure10}.a-\ref{fig:figure10}.d, except for a few minor differences. It is to be noted that for the maximum membrane nodal displacements, the variation with $f/L$ for $N_0$ = 8 kN/m and 15 kN/m are different from the TMS’s with only peripheral cables. Here the peak displacements decrease with $f/L$, which can be attributed to the prestress induced increase in membrane stiffness prevailing over the increase in pressure intensity. A comparison of the maximum structural responses with the key parameters for both cable configurations is presented in Table \ref{tab:comparison_responses}.  The values in the table are the highest peak response for both types of TMS (open and closed), spanning over all aerodynamic roughness (z0). For the deflection, there is a higher change with increasing prestress with only peripheral cables comparing the TMS with both radial and peripheral cables. The disparity margin increases with higher $f/L$ keeping the prestress identical. Although similar trends are observed for other responses, the percentage changes are much lower. \\
 
 Reduced deflections due to more cables also lead to enhanced principal stresses, cable tensions and support reactions, also observed in Table \ref{tab:comparison_responses}. Larger membrane stresses require higher membrane thickness, larger cable tensions, and higher reaction forces as design loads for the columns constituting the main wind force resisting system. This requires judicious assessment toward inclusion of radial cables in addition to peripheral ones for structural design of conical TMS.   

\begin{table}[htbp]
\centering
\caption{Comparison of maximum structural responses in TMS with only peripheral (P) and peripheral and radial (P+R) cables for varying key parameters. Values represent the maximum responses at the mid-node (Note: I = Increase in response, D = Decrease in response).}
\label{tab:comparison_responses}

\renewcommand{\arraystretch}{1.35}   
\setlength{\tabcolsep}{6pt}           

\resizebox{\textwidth}{!}{%
\begin{tabular}{|c|c|ccc|ccc|ccc|}
\hline
\multirow{2}{*}{Resp.} &
\multirow{2}{*}{No (kN/m)} &
\multicolumn{9}{c|}{$f/L$} \\
\cline{3-11}
 & & \multicolumn{3}{c|}{1/6} & \multicolumn{3}{c|}{1/3} & \multicolumn{3}{c|}{1/2} \\
\cline{3-11}
 & & P & P+R & \% diff. & P & P+R & \% diff. & P & P+R & \% diff. \\
\hline

\multirow{3}{*}{Deflection (m)}
 & 4  & 0.53 & 0.48 & 9.4 (D)  & 0.55 & 0.51 & 7.3 (D)  & 0.67 & 0.57 & 14.9 (D) \\
 & 8  & 0.48 & 0.47 & 2.0 (D)  & 0.50 & 0.40 & 20.0 (D) & 0.57 & 0.38 & 33.3 (D) \\
 & 15 & 0.40 & 0.33 & 17.5 (D) & 0.44 & 0.29 & 34.1 (D) & 0.52 & 0.28 & 46.2 (D) \\
\hline

\multirow{3}{*}{Stress (kN/m)}
 & 4  & 68.6 & 76.4 & 3.5 (I) & 71.5 & 73.5 & 2.8 (I) & 83.0 & 87.2 & 5.1 (I) \\
 & 8  & 76.1 & 79.6 & 4.6 (I) & 81.3 & 87.1 & 7.1 (I) & 89.8 & 92.5 & 2.7 (I) \\
 & 15 & 90.1 & 94.5 & 4.9 (I) & 96.7 & 108.8 & 12.5 (I) & 104.2 & 122.7 & 17.8 (I) \\
\hline

\multirow{3}{*}{Cable tension (kN)}
 & 4  & 194.6 & 209.1 & 7.2 (I) & 195.4 & 199.7 & 2.2 (I) & 219.5 & 219.5 & 0.0 (I) \\
 & 8  & 198.6 & 205.5 & 3.8 (I) & 205.9 & 218.2 & 6.4 (I) & 219.3 & 243.7 & 11.1 (I) \\
 & 15 & 258.0 & 272.2 & 5.5 (I) & 267.9 & 299.3 & 11.7 (I) & 268.0 & 313.4 & 17.0 (I) \\
\hline

\multirow{3}{*}{Support reaction (kN)}
 & 4  & 1260.6 & 1299.8 & 2.5 (I) & 1136.1 & 1155.8 & 1.7 (I) & 970.0 & 1007.3 & 3.8 (I) \\
 & 8  & 1380.0 & 1426.9 & 3.4 (I) & 1190.2 & 1165.7 & 5.9 (I) & 989.0 & 1085.6 & 10.8 (I) \\
 & 15 & 1385.7 & 1455.4 & 5.0 (I) & 1225.0 & 1362.8 & 11.3 (I) & 1085.9 & 1262.9 & 16.4 (I) \\
\hline
\end{tabular}
}

\end{table}
 
\section{Equivalent static loads for wind-resistant design}\label{sec:eq-static-loads}
Modern structural design for wind loads on buildings and other structures typically relies on the equivalent static wind load procedure (Simiu and Yeo, \cite{simiu2019wind}). In this approach, static wind loads are multiplied by a dynamic coefficient to account for the fluctuating nature of wind pressure. While effective for conventional structures, this method falls short when applied to membrane structures due to their inherent geometric nonlinearity.   \\

 In order to address this, a modified version of the equivalent static wind load procedure has been developed specifically for membrane structures, incorporating both a gust loading factor and an additional nonlinear adjustment factor (Kandel et al., \cite{kandel2021wind}; Chen et al., \cite{chen2022wind}). This approach mirrors the traditional method but with key modifications to better suit the unique characteristics of membranes. Chen et al. \cite{chen2022wind} introduced a refined expression for the gust loading factor, a critical component of this improved methodology as

\begin{equation}
\beta_{s_i} = 1 + \left\{ \mu_s \, \frac{\sigma_{s_i}}{\lvert \bar{s}_i \rvert} \right\}
\label{eq:beta_si}
\end{equation}

In this expression $\beta_{s_i}$ is the gust loading factor; $\mu_{s_i}$  is the peak factor;  $\bar{s}_i$ and $\sigma_{s_i}$  are the mean and standard deviation of any response, respectively; all of which are defined on i-th node of the TMS. \\

 In earlier studies on aerodynamic responses of TMS, researchers adopted rather a fixed peak factor value of 2.5 (Zhou et al., \cite{zhou2013research}). In spite of its convenience, this approach resulted in inconsistent assurance rates across different regions of TMS (Kandel et al., \cite{kandel2021wind}; Zhou et al., \cite{zhou2013research}). In this study, the gust loading factor is evaluated using the actual peak factor at each node on the TMS for achieving improved accuracy and realistic design.   Assigning the maximum gust loading factor across all nodes of TMS would be overly conservative. Instead, the gust loading factor for the TMS is calculated based on a refined method accounting for the variability of gust effects across the TMS as (Kandel et al., \cite{kandel2021wind}):   

 \begin{equation}
\beta_s^{*}
=
\left\{
\beta_{s_i}\,\times\,\bar{s}_i
\right\}_{\max}
\Big/
\left( \bar{s}_i \right)_{\max}
\label{eq:beta_star}
\end{equation}

where $\beta_s^{*}$  is a single “weighted” estimate of the gust loading factor for the whole TMS. This factor represents the ratio of the maximum dynamic response among all nodes to the maximum mean response across all the nodes. However, in actual practice, it should be the ratio of maximum dynamic response of all nodes to the maximum response of all nodes under mean static wind load. Highly geometric nonlinearity in TMS influences its static and dynamic behavior quite differently. Therefore, an additional nonlinear adjustment factor $\eta$  is defined as (Kandel et al., \cite{kandel2021wind}): 

\begin{equation}
\eta
=
\left\{ \bar{s}_i \right\}_{\max}
\Big/
\left\{ s_{st_i} \right\}_{\max}
\label{eq:eta}
\end{equation}

Symbol $s_{st_i}$  is the static response under mean wind. The equivalent static wind response  is then defined as   

\begin{equation}
s_{eq} = s_{st} \cdot \beta_s^{*} \cdot \eta
\label{eq:Seq}
\end{equation}

The proposal of the gust response factor in conjunction with the nonlinear adjustment factor therefore allows an equivalent static wind loads, bypassing the nonlinear dynamic analysis. The section below provides the gust loading factor and nonlinear adjustment factor for the TMS considered.

\subsection{Gust Response Factors (GRF)     }
GRFs for deflection and stress in the membrane are presented in this section for the conical TMS. Table \ref{tab:grf_disp_open} and Table \ref{tab:grf_disp_closed} show the values of GRFs for displacement of the open and closed TMS respectively with the key parameters (f/L, $z_0$ and $N_0$). Tables \ref{tab:grf_stress_open} and Table \ref{tab:grf_stress_closed} show the GRFs for membrane stress for the open and closed TMS, respectively. Considering the membrane displacements for both types of TMS, the GRFs tend to increase with $N_0$ under constant $z_0$ and $f/L$ of 1/6 to 1/3, however for $f/L$ of 1/2, they decrease up to $N_0$ = 8 kN/m followed by an increase. For the membrane stress however, no definite trend with any of the key parameters (f/L, $z_0$ and $N_0$) is observed. In both response quantities, the influence of membrane prestress is more significant than others. For constant GFR values, multi-linear regression models, proposed by Kandel et al. \cite{kandel2021wind} are fitted between the computed GRFs and the key parameters using the data in Tables \ref{tab:grf_disp_open}-\ref{tab:grf_stress_open}. The expressions for the regression model are presented in Table \ref{tab:grf_regression}.    

\begin{table*}[htbp]
\centering
\caption{GRFs for displacement for closed TMS}
\label{tab:grf_disp_closed}
\renewcommand{\arraystretch}{1.2}
\setlength{\tabcolsep}{6pt}

\begin{tabular}{c|ccc|ccc|ccc}
\hline
\multicolumn{10}{c}{\textbf{TMS with peripheral cables only}} \\
\hline
 & \multicolumn{3}{c|}{$f/L = 1/6$} &
   \multicolumn{3}{c|}{$f/L = 1/3$} &
   \multicolumn{3}{c}{$f/L = 1/2$} \\
\cline{2-4}\cline{5-7}\cline{8-10}
$z_0$ (cm)
& $N_0=4$ & $N_0=8$ & $N_0=15$
& $N_0=4$ & $N_0=8$ & $N_0=15$
& $N_0=4$ & $N_0=8$ & $N_0=15$ \\
\hline
0.1 & 1.41 & 1.65 & 2.00 & 1.48 & 1.72 & 2.00 & 1.57 & 1.72 & 1.97 \\
6   & 1.49 & 1.61 & 1.94 & 1.58 & 1.71 & 2.00 & 1.77 & 1.64 & 1.91 \\
80  & 1.56 & 1.64 & 1.77 & 1.63 & 1.66 & 1.79 & 1.75 & 1.57 & 1.75 \\
\hline
\multicolumn{10}{c}{\textbf{TMS with peripheral and radial cables}} \\
\hline
0.1 & 1.91 & 2.15 & 2.50 & 1.98 & 2.22 & 2.50 & 2.07 & 2.22 & 2.47 \\
6   & 1.99 & 2.11 & 2.44 & 2.08 & 2.21 & 2.50 & 2.27 & 2.14 & 2.41 \\
80  & 2.06 & 2.14 & 2.27 & 2.13 & 2.16 & 2.29 & 2.25 & 2.07 & 2.25 \\
\hline
\end{tabular}
\end{table*}



\begin{table}[htbp]
\centering
\caption{GRFs for displacement for open TMS}
\label{tab:grf_disp_open}
\renewcommand{\arraystretch}{1.2}
\setlength{\tabcolsep}{6pt}

\begin{tabular}{c|ccc|ccc|ccc}
\hline
\multicolumn{10}{c}{\textbf{TMS with peripheral cables only}} \\
\hline
$z_0$ (cm) &
\multicolumn{3}{c|}{$f/L = 1/6$} &
\multicolumn{3}{c|}{$f/L = 1/3$} &
\multicolumn{3}{c}{$f/L = 1/2$} \\
\cline{2-4}\cline{5-7}\cline{8-10}
& $N_0=4$ & $N_0=8$ & $N_0=15$
& $N_0=4$ & $N_0=8$ & $N_0=15$
& $N_0=4$ & $N_0=8$ & $N_0=15$ \\
\hline
0.1 & 1.45 & 1.62 & 1.97 & 1.51 & 1.66 & 1.96 & 1.68 & 1.63 & 1.95 \\
6   & 1.54 & 1.63 & 1.89 & 1.59 & 1.64 & 1.95 & 1.78 & 1.52 & 1.91 \\
80  & 1.61 & 1.62 & 1.74 & 1.62 & 1.67 & 1.80 & 1.76 & 1.58 & 1.76 \\
\hline
\multicolumn{10}{c}{\textbf{TMS with peripheral and radial cables}} \\
\hline
0.1 & 1.95 & 2.12 & 2.47 & 2.01 & 2.16 & 2.46 & 2.18 & 2.13 & 2.45 \\
6   & 2.04 & 2.13 & 2.39 & 2.09 & 2.14 & 2.45 & 2.28 & 2.02 & 2.41 \\
80  & 2.11 & 2.12 & 2.24 & 2.12 & 2.17 & 2.30 & 2.26 & 2.08 & 2.26 \\
\hline
\end{tabular}
\end{table}

\begin{table*}[!htbp]
\centering
\caption{GRFs for membrane stress for closed TMS}
\label{tab:grf_stress_closed}
\renewcommand{\arraystretch}{1.2}
\setlength{\tabcolsep}{6pt}

\begin{tabular}{c|ccc|ccc|ccc}
\hline
\multicolumn{10}{c}{\textbf{TMS with peripheral cables only}} \\
\hline
$z_0$ (cm)
& \multicolumn{3}{c|}{$f/L = 1/6$}
& \multicolumn{3}{c|}{$f/L = 1/3$}
& \multicolumn{3}{c}{$f/L = 1/2$} \\
\cline{2-4}\cline{5-7}\cline{8-10}
& $N_0=4$ & $N_0=8$ & $N_0=15$
& $N_0=4$ & $N_0=8$ & $N_0=15$
& $N_0=4$ & $N_0=8$ & $N_0=15$ \\
\hline
0.1 & 1.95 & 1.84 & 1.57 & 1.65 & 1.58 & 1.44 & 1.36 & 1.45 & 1.41 \\
6   & 1.89 & 1.91 & 1.84 & 1.69 & 1.79 & 1.70 & 1.47 & 1.56 & 1.49 \\
80  & 1.83 & 1.85 & 2.09 & 1.83 & 1.79 & 1.79 & 1.62 & 1.62 & 1.61 \\
\hline
\multicolumn{10}{c}{\textbf{TMS with peripheral and radial cables}} \\
\hline
0.1 & 2.31 & 2.16 & 2.00 & 2.03 & 1.92 & 1.88 & 1.75 & 1.81 & 1.82 \\
6   & 2.26 & 2.22 & 2.24 & 2.07 & 2.14 & 2.11 & 1.87 & 2.02 & 1.86 \\
80  & 2.22 & 2.21 & 2.33 & 2.09 & 2.13 & 2.13 & 1.98 & 1.98 & 1.96 \\
\hline
\end{tabular}
\end{table*}

\begin{table*}[!htbp]
\centering
\caption{GRFs for membrane stress for open TMS}
\label{tab:grf_stress_open}
\renewcommand{\arraystretch}{1.2}
\setlength{\tabcolsep}{6pt}

\begin{tabular}{c|ccc|ccc|ccc}
\hline
\multicolumn{10}{c}{\textbf{TMS with peripheral cables only}} \\
\hline
$z_0$ (cm)
& \multicolumn{3}{c|}{$f/L = 1/6$}
& \multicolumn{3}{c|}{$f/L = 1/3$}
& \multicolumn{3}{c}{$f/L = 1/2$} \\
\cline{2-4}\cline{5-7}\cline{8-10}
& $N_0=4$ & $N_0=8$ & $N_0=15$
& $N_0=4$ & $N_0=8$ & $N_0=15$
& $N_0=4$ & $N_0=8$ & $N_0=15$ \\
\hline
0.1 & 1.96 & 1.81 & 1.65 & 1.68 & 1.57 & 1.53 & 1.40 & 1.46 & 1.47 \\
6   & 1.91 & 1.87 & 1.89 & 1.72 & 1.79 & 1.76 & 1.52 & 1.67 & 1.51 \\
80  & 1.87 & 1.86 & 1.98 & 1.74 & 1.78 & 1.78 & 1.63 & 1.63 & 1.61 \\
\hline
\multicolumn{10}{c}{\textbf{TMS with peripheral and radial cables}} \\
\hline
0.1 & 2.30 & 2.19 & 1.92 & 1.99 & 1.93 & 1.79 & 1.71 & 1.80 & 1.76 \\
6   & 2.24 & 2.26 & 2.19 & 2.04 & 2.14 & 2.05 & 1.82 & 1.91 & 1.84 \\
80  & 2.18 & 2.20 & 2.44 & 2.18 & 2.14 & 2.14 & 1.97 & 1.97 & 1.96 \\
\hline
\end{tabular}
\end{table*}

\begin{table*}[htbp]
\centering
\caption{Multilinear regression models for GRFs}
\label{tab:grf_regression}
\renewcommand{\arraystretch}{1.3}

\begin{tabular}{l l l l}
\hline
Response type & Cable orientation & TMS type & GRF regression model \\
\hline
\multirow{4}{*}{Displacement}
& Peripheral only & Closed &
$\beta^{*}_{dc} = 1.40 - 0.30\frac{z_0}{h} + 0.19\frac{f}{L} + 32.93\frac{N_0}{Et}$ \\
& Peripheral only & Open &
$\beta^{*}_{do} = 1.43 - 0.19\frac{z_0}{h} + 0.17\frac{f}{L} + 28.34\frac{N_0}{Et}$ \\
& Radial \& peripheral & Closed &
$\beta^{*}_{dc} = 1.89 - 0.22\frac{z_0}{h} + 0.23\frac{f}{L} + 32.13\frac{N_0}{Et}$ \\
& Radial \& peripheral & Open &
$\beta^{*}_{do} = 1.91 - 0.18\frac{z_0}{h} + 0.23\frac{f}{L} + 28.58\frac{N_0}{Et}$ \\
\hline
\multirow{4}{*}{Membrane stress}
& Peripheral only & Closed &
$\beta^{*}_{sc} = 2.02 + 0.86\frac{z_0}{h} - 1.06\frac{f}{L} - 4.28\frac{N_0}{Et}$ \\
& Peripheral only & Open &
$\beta^{*}_{so} = 2.02 + 0.56\frac{z_0}{h} - 0.97\frac{f}{L} - 2.81\frac{N_0}{Et}$ \\
& Radial \& peripheral & Closed &
$\beta^{*}_{sc} = 2.36 + 0.93\frac{z_0}{h} - 1.03\frac{f}{L} - 4.93\frac{N_0}{Et}$ \\
& Radial \& peripheral & Open &
$\beta^{*}_{so} = 2.35 + 0.57\frac{z_0}{h} - 0.93\frac{f}{L} - 3.35\frac{N_0}{Et}$ \\
\hline
\end{tabular}
\end{table*}

\subsection{Nonlinear Adjustment Factors (NAF)    }

Nonlinear adjustment factors for displacement and membrane stress for the conical TMS with various key parameters are presented in Tables \ref{tab:disp_closed}-\ref{tab:stress_open}. Like the GRF, similar trends are observed for nonlinear adjustment factors as well with respect to the key parameters except for the membrane displacements for the open TMS with radial cables, the NAFs decrease with increasing $N_0$ under constant $f/L$ and $z_0$. Interestingly, for the membrane stress, the NAFs increase for closed TMS and decrease for the open with $N_0$ under constant $f/L$ and $z_0$. Once again, the NAFs for the stresses are greater than those of the displacements. It is also noteworthy that the NAFs show less variation than GRFs. A similar multilinear regression model is fitted to the NAF analogous to the GRF based on the data presented in Tables \ref{tab:disp_closed}-\ref{tab:stress_open}. Table \ref{tab:regression} shows the regression models for the NAFs for displacement and stress.  \\

\begin{table}[htbp]
\centering
\caption{NAFs for displacement for closed TMS}
\label{tab:disp_closed}
\renewcommand{\arraystretch}{1.2}
\begin{tabular}{c|ccc|ccc|ccc}
\hline
 & \multicolumn{9}{c}{TMS with peripheral cables only} \\
\hline
 & \multicolumn{3}{c|}{$f/L=1/6$} 
 & \multicolumn{3}{c|}{$f/L=1/3$} 
 & \multicolumn{3}{c}{$f/L=1/2$} \\
\cline{2-4}\cline{5-7}\cline{8-10}
$z_0$ (cm) 
& $N_0=4$ & $N_0=8$ & $N_0=15$ 
& $N_0=4$ & $N_0=8$ & $N_0=15$ 
& $N_0=4$ & $N_0=8$ & $N_0=15$ \\
\hline
0.1 & 1.17 & 1.08 & 1.20 & 1.13 & 1.07 & 1.15 & 1.09 & 1.09 & 1.11 \\
6   & 1.26 & 1.24 & 1.36 & 1.21 & 1.22 & 1.33 & 1.13 & 1.22 & 1.25 \\
80  & 1.39 & 1.51 & 1.76 & 1.35 & 1.53 & 1.74 & 1.31 & 1.52 & 1.56 \\
\hline
\multicolumn{10}{c}{TMS with peripheral and radial cables} \\
\hline
0.1 & 1.21 & 1.12 & 1.24 & 1.17 & 1.11 & 1.19 & 1.13 & 1.13 & 1.15 \\
6   & 1.30 & 1.28 & 1.40 & 1.25 & 1.26 & 1.37 & 1.17 & 1.26 & 1.29 \\
80  & 1.43 & 1.55 & 1.80 & 1.39 & 1.57 & 1.78 & 1.35 & 1.56 & 1.60 \\
\hline
\end{tabular}
\end{table}

\begin{table}[htbp]
\centering
\caption{NAFs for displacement for open TMS}
\label{tab:disp_open}
\renewcommand{\arraystretch}{1.2}
\begin{tabular}{c|ccc|ccc|ccc}
\hline
 & \multicolumn{9}{c}{TMS with peripheral cables only} \\
\hline
 & \multicolumn{3}{c|}{$f/L=1/6$} 
 & \multicolumn{3}{c|}{$f/L=1/3$} 
 & \multicolumn{3}{c}{$f/L=1/2$} \\
\cline{2-4}\cline{5-7}\cline{8-10}
$z_0$ (cm) 
& $N_0=4$ & $N_0=8$ & $N_0=15$ 
& $N_0=4$ & $N_0=8$ & $N_0=15$ 
& $N_0=4$ & $N_0=8$ & $N_0=15$ \\
\hline
0.1 & 1.10 & 1.00 & 0.92 & 1.05 & 0.99 & 0.82 & 0.97 & 1.06 & 0.78 \\
6   & 1.17 & 1.14 & 1.06 & 1.11 & 1.14 & 0.95 & 1.00 & 1.22 & 0.87 \\
80  & 1.32 & 1.38 & 1.35 & 1.28 & 1.36 & 1.23 & 1.19 & 1.41 & 1.09 \\
\hline
\multicolumn{10}{c}{TMS with peripheral and radial cables} \\
\hline
0.1 & 1.14 & 1.04 & 0.96 & 1.09 & 1.03 & 0.86 & 1.01 & 1.10 & 0.82 \\
6   & 1.21 & 1.18 & 1.10 & 1.15 & 1.18 & 0.99 & 1.04 & 1.26 & 0.91 \\
80  & 1.36 & 1.42 & 1.39 & 1.32 & 1.40 & 1.27 & 1.23 & 1.45 & 1.13 \\
\hline
\end{tabular}
\end{table}

\begin{table}[htbp]
\centering
\caption{NAFs for membrane stress for closed TMS}
\label{tab:stress_closed}
\renewcommand{\arraystretch}{1.2}
\begin{tabular}{c|ccc|ccc|ccc}
\hline
 & \multicolumn{9}{c}{TMS with peripheral cables only} \\
\hline
 & \multicolumn{3}{c|}{$f/L=1/6$} 
 & \multicolumn{3}{c|}{$f/L=1/3$} 
 & \multicolumn{3}{c}{$f/L=1/2$} \\
\cline{2-4}\cline{5-7}\cline{8-10}
$z_0$ (cm) 
& $N_0=4$ & $N_0=8$ & $N_0=15$ 
& $N_0=4$ & $N_0=8$ & $N_0=15$ 
& $N_0=4$ & $N_0=8$ & $N_0=15$ \\
\hline
0.1 & 1.03 & 1.10 & 1.11 & 1.02 & 1.02 & 1.05 & 1.01 & 1.01 & 1.05 \\
6   & 1.24 & 1.31 & 1.31 & 1.16 & 1.13 & 1.14 & 1.08 & 1.08 & 1.17 \\
80  & 1.46 & 1.76 & 1.66 & 1.27 & 1.39 & 1.41 & 1.15 & 1.20 & 1.28 \\
\hline
\multicolumn{10}{c}{TMS with peripheral and radial cables} \\
\hline
0.1 & 1.07 & 1.14 & 1.15 & 1.06 & 1.06 & 1.09 & 1.05 & 1.05 & 1.09 \\
6   & 1.28 & 1.35 & 1.35 & 1.20 & 1.17 & 1.18 & 1.12 & 1.12 & 1.21 \\
80  & 1.50 & 1.80 & 1.70 & 1.31 & 1.43 & 1.45 & 1.19 & 1.24 & 1.32 \\
\hline
\end{tabular}
\end{table}

\begin{table}[htbp]
\centering
\caption{NAFs for membrane stress for open TMS}
\label{tab:stress_open}
\renewcommand{\arraystretch}{1.2}
\begin{tabular}{c|ccc|ccc|ccc}
\hline
 & \multicolumn{9}{c}{TMS with peripheral cables only} \\
\hline
 & \multicolumn{3}{c|}{$f/L=1/6$} 
 & \multicolumn{3}{c|}{$f/L=1/3$} 
 & \multicolumn{3}{c}{$f/L=1/2$} \\
\cline{2-4}\cline{5-7}\cline{8-10}
$z_0$ (cm) 
& $N_0=4$ & $N_0=8$ & $N_0=15$ 
& $N_0=4$ & $N_0=8$ & $N_0=15$ 
& $N_0=4$ & $N_0=8$ & $N_0=15$ \\
\hline
0.1 & 0.98 & 0.94 & 0.82 & 0.93 & 0.86 & 0.83 & 0.98 & 0.88 & 0.87 \\
6   & 1.16 & 1.14 & 0.97 & 1.07 & 0.96 & 0.90 & 1.03 & 0.92 & 0.98 \\
80  & 1.37 & 1.48 & 1.23 & 1.22 & 1.18 & 1.11 & 1.11 & 1.04 & 1.08 \\
\hline
\multicolumn{10}{c}{TMS with peripheral and radial cables} \\
\hline
0.1 & 1.02 & 0.98 & 0.86 & 0.97 & 0.90 & 0.87 & 1.02 & 0.92 & 0.91 \\
6   & 1.20 & 1.18 & 1.01 & 1.11 & 1.00 & 0.94 & 1.07 & 0.96 & 1.02 \\
80  & 1.41 & 1.52 & 1.27 & 1.26 & 1.22 & 1.15 & 1.15 & 1.12 & 1.08 \\
\hline
\end{tabular}
\end{table}

\begin{table}[htbp]
\centering
\caption{Regression model for the NAFs}
\label{tab:regression}
\renewcommand{\arraystretch}{1.4}
\begin{tabular}{c c c l}
\hline
Response type & Cable orientation & TMS type & Regression model for the NAFs \\
\hline
Displacement & Peripheral only & Closed &
$\eta_{dc}=1.12+2.04\frac{z_0}{h}-0.24\frac{f}{L}+15.91\frac{N_0}{Et}$ \\
& & Open &
$\eta_{do}=1.43-0.19\frac{z_0}{h}+0.17\frac{f}{L}+28.34\frac{N_0}{Et}$ \\

& Radial and peripheral & Closed &
$\eta_{dc}=1.43+2.04\frac{z_0}{h}-0.23\frac{f}{L}-17.33\frac{N_0}{Et}$ \\
& & Open &
$\eta_{do}=1.31+1.64\frac{z_0}{h}-0.29\frac{f}{L}-20.09\frac{N_0}{Et}$ \\
\hline
Membrane stress & Peripheral only & Closed &
$\eta_{sc}=1.25+1.75\frac{z_0}{h}-0.65\frac{f}{L}+7.93\frac{N_0}{Et}$ \\
& & Open &
$\eta_{so}=2.02+0.56\frac{z_0}{h}-0.97\frac{f}{L}-2.81\frac{N_0}{Et}$ \\

& Radial and peripheral & Closed &
$\eta_{sc}=1.43+1.75\frac{z_0}{h}-0.65\frac{f}{L}-9.42\frac{N_0}{Et}$ \\
& & Open &
$\eta_{so}=1.23+1.50\frac{z_0}{h}-0.40\frac{f}{L}-13.9\frac{N_0}{Et}$ \\
\hline
\end{tabular}
\end{table}

\begin{figure}[htbp!]
    \centering
    \hspace*{-.75cm}
    \includegraphics[width=.82\textwidth]{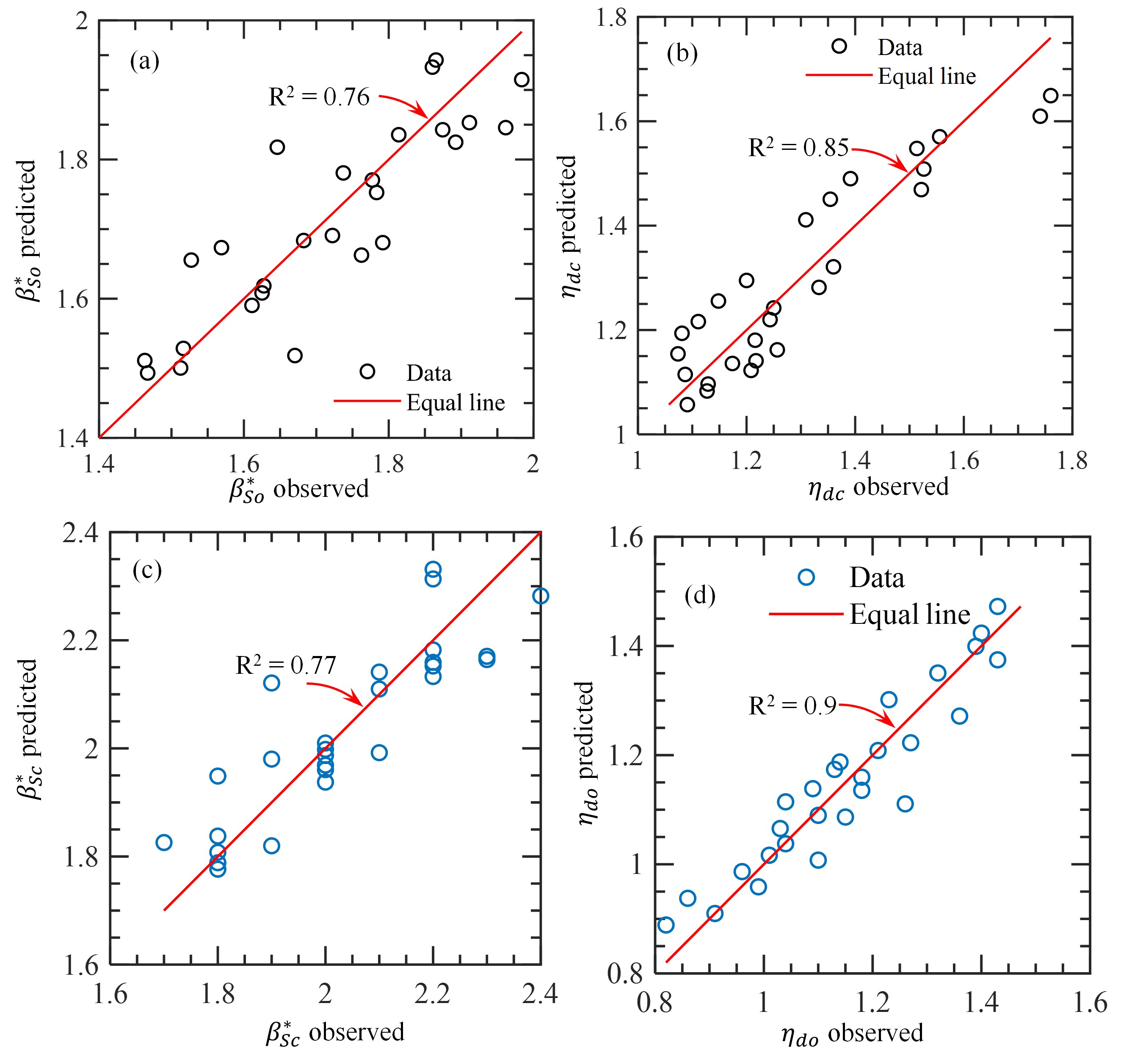}
    \caption{
    Goodness of fit plots for the (a) GRFs for the stress for open TMS, (b) NAFs for displacement in the closed TMS with peripheral cables only and; (c) GRFs for stress for closed TMS,  (d) NAFs for displacement for the open TMS with radial and peripheral cables   }
    \label{fig:figure12}
\end{figure}

 The legitimacy of the regression models for the GRF and the NAFs are checked by the goodness of fit tests. The predicted values of the GRFs and the NAFs by the regression models are compared with their observed values of the displacements for the open and closed TMS in Fig. \ref{fig:figure12}.a and Fig. \ref{fig:figure12}.b, respectively, considering only the peripheral cables. Figures 12.c and 12.d show the same for the open TMS with both radial and peripheral cables. The respective R2 values for these regression models are estimated to be in acceptable range to justify the viability of the regression models. 

 \subsection{Probabilistic assessment of the GRFs and NAFs  }

\begin{figure}[htbp!]
    \centering
    \hspace*{-.75cm}
    \includegraphics[width=.82\textwidth]{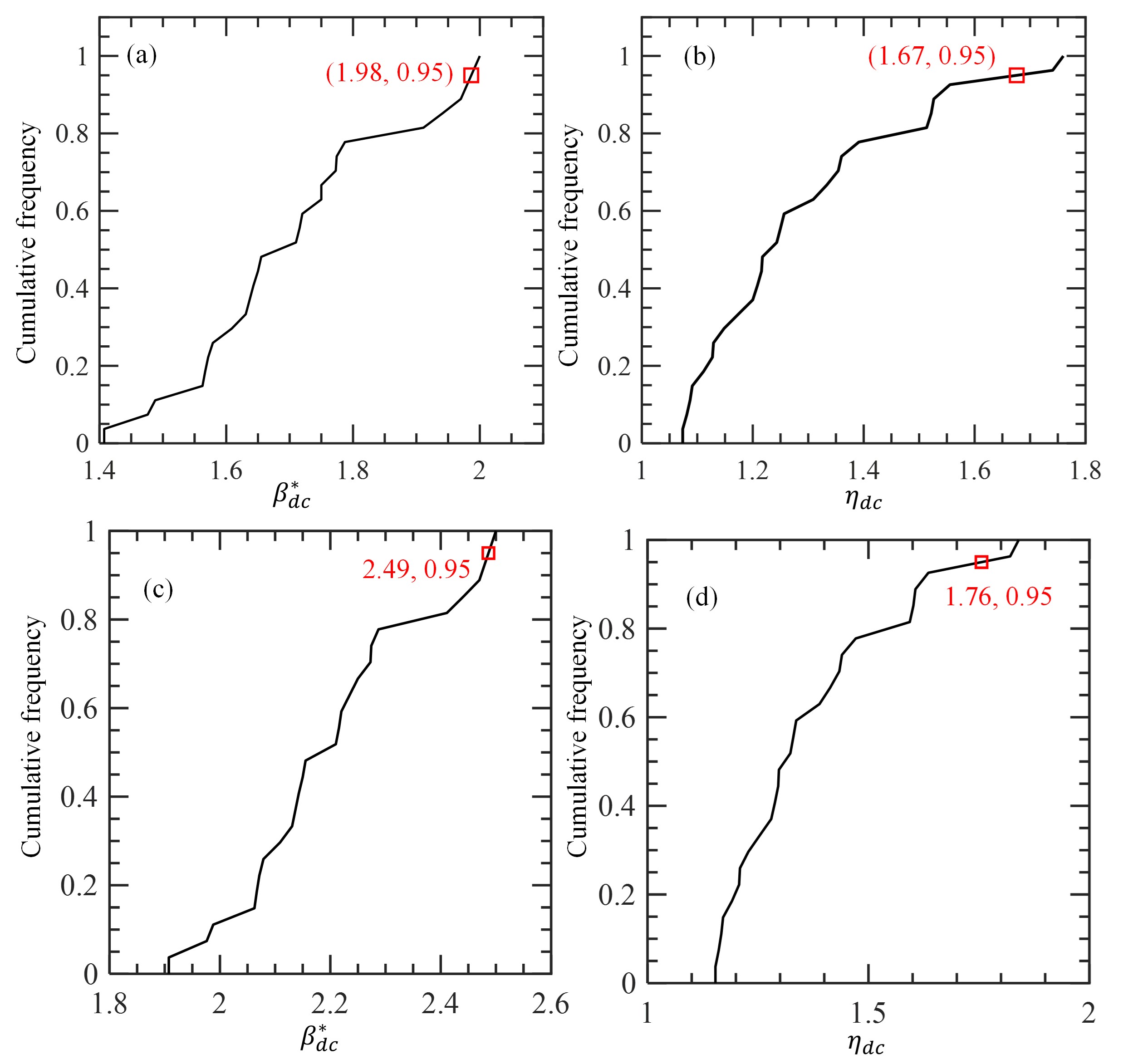}
    \caption{
    Cumulative distribution frequency for the (a) GRFs, (b) NAFs for the deflection for closed TMS with only peripheral cables and; (c) GRFs, (d) NAFs for displacement for closed TMS with radial and peripheral cables    }
    \label{fig:figure13}
\end{figure}

 Assigning a single GRF and NAF for the entire TMS is more efficient rather than accounting its variations across the TMS surface. Such design GRFs and NAFs are derived from the cumulative distribution function (CDF) of the pertinent response variables, treating the GRFs as random. Fig. \ref{fig:figure13}.a and \ref{fig:figure13}.b show the CDFs for the GRF and the NAFs for the deflection in TMS with only peripheral cables. The same for the deflection in the open and closed TMS with both radial and peripheral cables are presented in Fig. \ref{fig:figure13}.c and \ref{fig:figure13}.d, respectively. The GRF and the NAF values corresponding to their 95th percentile cumulative frequency are then chosen to be the probabilistic design values by ensuring a balance between realism (i.e. the scatter among the GRFs) and also by avoiding the excessive conservatism. A summary of these probabilistic design values is also provided in Table \ref{tab:grf_naf_p95} for both the displacement and the membrane stress in enclosed/open TMS with only peripheral cables and with radial and peripheral cables.

\begin{table}[htbp]
\centering
\caption{95th percentile values of the GRFs and NAFs for TMS}
\label{tab:grf_naf_p95}
\renewcommand{\arraystretch}{1.2}
\setlength{\tabcolsep}{8pt}

\begin{tabular}{llcccc}
\hline
\multirow{2}{*}{TMS type} & \multirow{2}{*}{Response} &
\multicolumn{2}{c}{\textbf{TMS with peripheral cables only}} &
\multicolumn{2}{c}{\textbf{TMS with peripheral and radial cables}} \\
\cline{3-6}
& & GRFs $(\beta^{*})$ & NAFs $(\eta)$ & GRFs $(\beta^{*})$ & NAFs $(\eta)$ \\
\hline
\multirow{2}{*}{Closed}
& Displacement      & 1.98 & 1.67 & 2.49 & 1.41 \\
& Membrane stress   & 1.93 & 1.59 & 2.44 & 1.67 \\
\multirow{2}{*}{Open}
& Displacement      & 1.95 & 1.95 & 2.46 & 1.45 \\
& Membrane stress   & 1.94 & 1.94 & 2.44 & 1.41 \\
\hline
\end{tabular}
\end{table}

\section{Conclusions}\label{sec:conclusion}

This work presents an LES based evaluation of wind pressure and resulting nonlinear aerodynamic response of conical TMS, with particular emphasis on the influence of varying key parameters (rise-span ratio, $f/L$; aerodynamic roughness height, $z_0$ and membrane prestress, $N_0$) as well as the location and orientation of the membrane supporting cables. The influence of key parameters on the wind-induced responses of the TMS are presented with an equivalent static wind resistant design. The proposed wind resistant design relies on the key concept of GRFs, on which additional NAFs are employed, accounting for geometric nonlinearity. The investigation reveals the following major points

\begin{itemize}
  \item Increasing prestress and rise-to-span ratio significantly influence the
  stiffness of the TMS and hence the increased modal frequency. Prestress is
  found to be more dominant than the rise-to-span ratio. A monotonic increase in
  all pertinent responses is observed with increasing aerodynamic roughness,
  attributed to increased peak factors for fluctuating pressure coefficients
  caused by enhanced turbulence intensity.

  \item Contours of maximum in-plane membrane principal stresses and
  displacements show larger values for TMS with both radial and peripheral
  cables, compared to TMS with only peripheral cables. Open TMS experiences
  higher responses than closed TMS. Maximum stresses are concentrated at the
  leeward side of the open TMS and on either side of the flow direction for the
  closed configuration.

  \item Wind-induced responses generally increase with rise-to-span ratio, except
  for reaction forces at the base of the TMS. Increasing membrane prestress leads
  to increased responses except for deflection. Peak displacements corresponding
  to higher prestress levels of 8~kN/m and 15~kN/m decrease with increasing $f/L$
  ratios for TMS with both radial and peripheral cables, unlike TMS with only
  peripheral cables.

  \item Consistent with all peak structural responses, the percentage increase in
  response values for TMS with only peripheral cables (relative to TMS with both
  radial and peripheral cables) is lower, except for displacement. For peak
  displacements, increasing prestress produces a greater change for TMS with
  only peripheral cables.

  \item Influence of key parameters on the GRFs and NAFs are presented. The effect of the membrane prestress on the GRFs is more significant than the other parameters. Multi-linear regression models are fitted for the GRFs and NAFs for predictive modeling. For convenient design practice, their percentile values are provided treating their spatial distribution on the TMS surface as random.      
\end{itemize}

This study takes the first step towards comprehensive nonlinear structural response analysis of an anticlastic conical TMS and the evaluation of an equivalent static wind load design considering various key parameters. It is anticipated that the proposed GRFs and the NAFS would be useful for robust yet economic designs. Nevertheless, this study may be complimented with the wind tunnel experiments on scaled TMS.  

\section*{Acknowledgments}
The authors would like to acknowledge the Ministry of Human Resources department (MHRD), Government of India for providing the financial resources in support of the present research.

\bibliographystyle{unsrt}  
\bibliography{biblio.bib} 

\end{document}